\documentclass[]{aastex631}

\def\HI{\hbox{H{\sc i}}}

\def\km{KM3--230213A}
\def\ugca{UGCA~127}
\def\ugcb{Phaedra}
\def\wisegala{WISEA~J061715.89$-$075455.4}
\def\wisegalb{Hebe}
\def\name{$\phi\alpha i\delta\rho\alpha$}
\def\bright{$\varphi\alpha\iota\delta\rho o\varsigma$}
\def\nametwo{H$\beta\eta$}
\def\namethree{Narcissus}
\def\namethreeb{$N\alpha\rho\kappa\iota\sigma\sigma o \zeta $} %

\def\totnumps{1045}
\def\totnum{1052}

\definecolor{darkbrown}{rgb}{0.4, 0.26, 0.13}

\usepackage{ulem}
\usepackage{comment}
\usepackage{soul}
\usepackage{graphicx,booktabs}

\usepackage[nolist]{acronym}
\begin{acronym}[AWGN]
\acro{AGN}{active galactic nucleus}
\acro{ASKAP}{Australian Square Kilometre Array Pathfinder}
\acro{CASDA}{CSIRO ASKAP Science Data Archive}
\acro{CMB}{cosmic microwave background}
\acro{CR}{cosmic ray}
\acro{CSIRO}{Australian Commonwealth Scientific and Industrial Research Organisation}
\acro{EMU}{Evolutionary Map of the Universe}
\acro{Gaia DR3}{{\it Gaia} Data Release 3}
\acro{GLEAM-X}{GaLactic and Extragalactic All-sky MWA eXtended}
\acro{GLEAM}{GaLactic and Extragalactic All-sky MWA}
\acro{GPS}{gigahertz peaked spectrum}
\acro{GRB}{gamma-ray burst}
\acrodefplural{LIRG}{luminous IR galaxies}
\acro{LIRG}{luminous IR galaxy}
\acro{NVSS}{NRAO VLA Sky Survey}
\acro{PMN}{Parkes-MIT-NRAO}
\acro{POSSUM}{Polarization Sky Survey of the Universe's Magnetism}
\acro{RACS}{Rapid ASKAP Continuum Survey}
\acro{RM}{Rotation Measure}
\acro{RMS}{root mean squared}
\acro{RNe}{Reflection Nebula}
\acro{SFR}{star formation rate}
\acro{SN}{supernova}
\acro{SNR}{supernova remnant}
\acro{SMBH}{super massive black hole}
\acro{UHE}{ultra-high-energy}
\acro{VAST}{Variable and Slow Transients}
\acro{VLASS}{Very Large Array Sky Survey}
\acro{VLBA}{Very Long Baseline Array}
\acro{VLBI}{Very Long Baseline Interferometry}
\acro{XRB}{X-ray binary}
\end{acronym}

\begin{document}

\title{ASKAP and VLASS search for a radio-continuum counterpart of ultra-high-energy neutrino event KM3--230213A}

\correspondingauthor{M. D. Filipovi\'c}
\email{m.filipovic@westernsydney.edu.au}

\author[0000-0002-4990-9288]{M. D. Filipovi\'c}
\affiliation{Western Sydney University, Locked Bag 1797, Penrith South DC, NSW 2751, Australia}

\author[0009-0009-7061-0553]{Z. J. Smeaton}
\affiliation{Western Sydney University, Locked Bag 1797, Penrith South DC, NSW 2751, Australia}

\author[0009-0007-4802-2793]{A. C. Bradley}
\affiliation{Western Sydney University, Locked Bag 1797, Penrith South DC, NSW 2751, Australia}

\author[0000-0003-0699-7019]{D. Dobie}
\affiliation{Sydney Institute for Astronomy, School of Physics, University of Sydney, Sydney, NSW 2006, Australia}
\affiliation{ARC Centre of Excellence for Gravitational Wave Discovery (OzGrav), Hawthorn, Victoria, 3122, Australia}

\author[0000-0003-4351-993X]{B. S. Koribalski}
\affiliation{Australia Telescope National Facility, CSIRO, Space and Astronomy, P.O. Box 76, Epping, NSW 1710, Australia}
\affiliation{Western Sydney University, Locked Bag 1797, Penrith South DC, NSW 2751, Australia} 

\author[0000-0001-5953-0100]{R. Kothes}
\affiliation{Dominion Radio Astrophysical Observatory, Herzberg Astronomy \& Astrophysics, National Research Council Canada, P.O. Box 248, Penticton}

\author[0000-0001-5636-7213]{L. Rudnick}
\affiliation{Minnesota Institute for Astrophysics, University of Minnesota, Minneapolis, MN, 55455, USA}

\author[0000-0002-0457-3661]{A. Ahmad}
\affiliation{Western Sydney University, Locked Bag 1797, Penrith South DC, NSW 2751, Australia}

\author[0000-0001-5609-7372]{R. Z. E. Alsaberi}
\affiliation{Western Sydney University, Locked Bag 1797, Penrith South DC, NSW 2751, Australia}
\affiliation{Faculty of Engineering, Gifu University, 1-1 Yanagido, Gifu 501-1193, Japan}

\author[0000-0002-6243-7879]{C. S. Anderson}
\affiliation{Research School of Astronomy \& Astrophysics, The Australian National University, Canberra ACT 2611, Australia}

\author[0000-0002-0016-9485]{L. A. Barnes}
\affiliation{Western Sydney University, Locked Bag 1797, Penrith South DC, NSW 2751, Australia}

\author[0000-0003-0268-5122]{M. Breuhaus}
\affiliation{Centre de Physique des Particules de Marseille, 163, avenue de Luminy, 13009 Marseille, France}

\author[0000-0001-5197-1091]{E. J. Crawford}
\affiliation{Western Sydney University, Locked Bag 1797, Penrith South DC, NSW 2751, Australia}

\author[0000-0002-9618-2499]{S. Dai}
\affiliation{Australia Telescope National Facility, CSIRO, Space and Astronomy, P.O. Box 76, Epping, NSW 1710, Australia}
\affiliation{Western Sydney University, Locked Bag 1797, Penrith South DC, NSW 2751, Australia} 

\author[0000-0003-1432-253X]{Y. A. Gordon}
\affiliation{Department of Physics, University of Wisconsin-Madison, 1150 University Avenue, Madison, WI 53706, USA}

\author[0000-0001-7652-9451]{N. Gupta}
\affiliation{CSIRO Space \& Astronomy, Bentley, WA, Australia}

\author[0000-0002-6097-2747]{A. M. Hopkins}
\affiliation{School of Mathematical and Physical Sciences, 12 Wally's Walk, Macquarie University, NSW 2109, Australia}

\author[0000-0002-4814-958X]{D. Leahy}
\affiliation{Department of Physics and Astronomy, University of Calgary, Calgary, Alberta, T2N IN4, Canada}

\author[0000-0002-6147-693X]{K. J. Luken}
\affiliation{Western Sydney University, Locked Bag 1797, Penrith South DC, NSW 2751, Australia}

\author[0000-0003-2730-957X]{N. McClure-Griffiths}
\affiliation{Research School of Astronomy \& Astrophysics, The Australian National University, Canberra ACT 2611, Australia}

\author[0000-0001-9033-4140]{M. J. Michałowski}
\affiliation{Astronomical Observatory Institute, Faculty of Physics and Astronomy, Adam Mickiewicz University, ul. Słoneczna 36, 60-286 Poznań, Poland}

\author[0000-0001-5302-1866]{M. Sasaki}
\affiliation{Dr Karl Remeis Observatory, Erlangen Centre for Astroparticle Physics, Friedrich-Alexander-Universit\"{a}t Erlangen-N\"{u}rnberg, Sternwartstra{\ss}e 7, 96049 Bamberg, Germany}

\author[0000-0002-9931-5162]{N. F. H. Tothill}
\affiliation{Western Sydney University, Locked Bag 1797, Penrith South DC, NSW 2751, Australia}

\author[0000-0002-6972-8388]{G. M. Umana}
\affiliation{INAF - Osservatorio Astrofisico di Catania, Via Santa Sofia 78, 95123 Catania, Italy}

\author[0000-0001-7093-3875]{T. Vernstrom}
\affiliation{Australia Telescope National Facility, CSIRO, Space and Astronomy, PO Box 1130, Bentley, WA 6102, Australia}

\author[0000-0001-7722-8458]{J. West}
\affiliation{Department of Physics and Astronomy, University of Calgary, Calgary, Alberta, T2N IN4, Canada}

\begin{abstract}

We present the results of an \ac{ASKAP} 944~MHz and \ac{VLASS} 3~GHz search for a radio-continuum counterpart of the recent \ac{UHE} neutrino event, \km. Using \ac{ASKAP}, we catalog \totnum\ radio sources within the 1\fdg5 radius search area (68\% certainty region) around the particle's calculated origin, 10 of which we classify as blazar candidates based on their radio spectra. The most prominent radio source in the search area is the nearby spiral galaxy \ugca\ (nicknamed Phaedra\footnote{From Greek: \name, a Cretan princess of Greek Mythology, derived from Phaidros, Greek: \bright, meaning 'bright'.}). Its non-thermal radio spectrum classifies it as a non-blazar \ac{AGN}. We also present an extended radio source, \wisegala\ (nicknamed Hebe\footnote{From Greek: \nametwo, the Greek goddess of youth.}), located only $\sim$7\arcmin\ from the geometric center of the search area,  with a very unusual highly polarized compact component. Finally, we present a strong radio source, EMU~J062248$-$072246 (nicknamed \namethree\footnote{From Greek \namethreeb\ was a self-absorbed hunter from Thespiae in Boeotia.}), which can be modeled either with a synchrotron self-absorbed spectrum or synchrotron emission from thermal electrons. It exhibits $\sim25\%$ flux density variability over the $\sim$5-year \ac{VLASS} 3~GHz survey.

\end{abstract}


\section{Introduction} 
 \label{sec:intro}

Currently, only three astrophysical sources are confirmed as neutrino emitters: our Sun, the supernova SN~1987A, and the Galactic plane \citep{book2, 2023Sci...380.1338I}. Other very likely neutrino emitters are the blazar TXS~0506+056 and the active galaxy NGC\,1068~\citep{2022Sci...378..538I}. Other possible neutrino origins include cosmogenic, Galactic, and extragalactic sources. Astrophysical neutrinos above GeV energies are typically produced by high-energy \acp{CR} interacting with matter or lower-energy photons, and these \acp{CR} originate from a variety of sources such as \acp{SNR}, \acp{SN}, microquasars, star-forming galaxies, \acp{GRB}, \acp{SMBH}, and \acp{AGN} \citep{book2}. The observed and theorized neutrino energies differ between these sources, with \acp{SN} at the lower energies~\citep[$<$100\,MeV;][]{1987PhRvL..58.1490H}, followed by star-forming galaxies \citep[100\,GeV up to PeV;][]{2015MNRAS.453..113B, 2020ApJ...894..112F}, \acp{SNR} \citep[up to PeV or higher when embedded in dense star clusters;][]{2015MNRAS.453..113B, 2023MNRAS.519..136V}, extending up to \acp{AGN}~\citep[up to EeV;][]{2021PhRvL.126s1101R}.

The recent KM3Net discovery on 13$^{\rm th}$ February 2025 of the \ac{UHE} neutrino \citep[KM3--230213A, hereafter \km;][]{KM3nat} provoked a worldwide search for its origin. The \ac{UHE} neutrino's origin was pinpointed to within 3\arcdeg\ of RA(J2000)\,=\,94\fdg3, Dec(J2000)\,=\,$-$7\fdg8 with a 99\% certainty, and to within 1\fdg5 with a 68\% certainty. Being close to the equator, it is well-positioned for analysis by numerous radio surveys, including the Evolutionary Map of the Universe~\citep[EMU;][Hopkins~et~al.,~submitted]{Norris2021}\acused{EMU} and the Polarization Sky Survey of the Universe's Magnetism~\citep[POSSUM;][Gaensler et al., submitted]{Gaensler2010}\acused{POSSUM} surveys, both conducted with \ac{ASKAP}~\citep{2021PASA...38....9H}, as well as the multi-epoch \ac{VLASS}~\citep{2020PASP..132c5001L} survey. 

Several origins for \km\ have been investigated, including Galactic, which was deemed unlikely \citep{2025arXiv250208387A}, a \ac{GRB} origin~\citep{grb}, a cosmogenic origin~\citep{2025arXiv250208508T, 2025arXiv250208484K}, and a blazar origin, specifically the well-known blazar, PMN\,J0606--0724, which experienced a radio flare within five days of the \km\ event~\citep{2025arXiv250208484K}. We here present the results of a search for a potential radio-continuum counterpart of the \km\ event using the \ac{ASKAP}-\ac{EMU} survey and the Quick-Look images from \ac{VLASS} \citep{2020PASP..132c5001L}. The data used in this paper are explained in Appendix~\ref{sec:data}.

\begin{figure*}[ht!]
    \centering
    \includegraphics[width=1\linewidth]{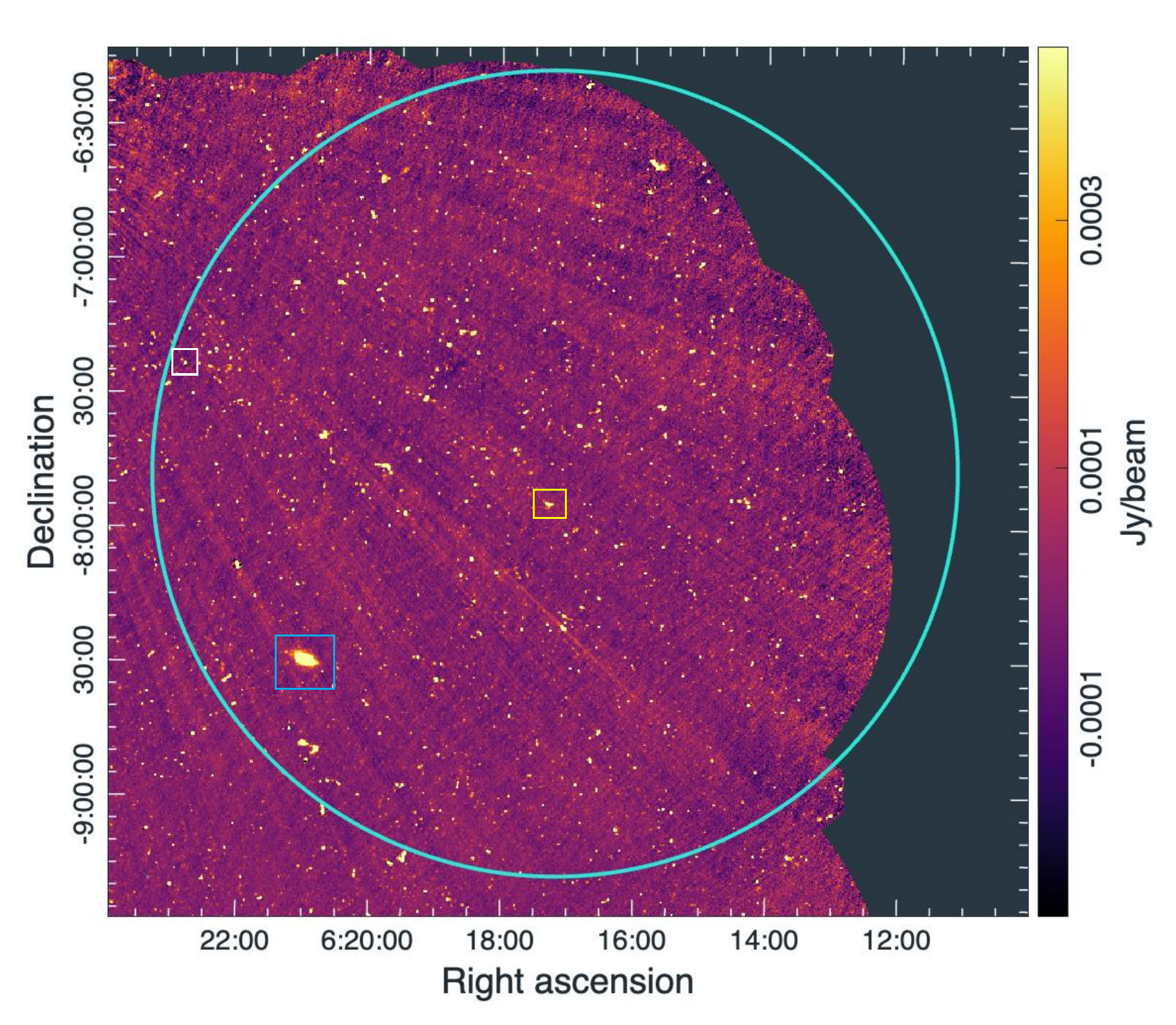}
\caption{ASKAP 944\,MHz radio continuum image, SB~59692. The cyan-colored circle (radius = 1\fdg5) indicates the 68\% confidence area of the neutrino event. The image is linearly scaled and has a synthesized beam of 15\arcsec. Colored squares show the locations of three objects of interest; the white square shows the variable source \namethree\ (EMU~J062244--072333; see Sec.~\ref{subsubsec:VLASS}), the yellow square shows \wisegalb\ (\wisegala; see Sec.~\ref{subsubsec:wisegal_Discussion}), and the blue square shows \ugcb\ (\ugca; see Sec.~\ref{subsubsec:ugc_Discussion}).}
    \label{fig:fig1}
\end{figure*}

\begin{figure*}
    \centering
    \includegraphics[width=1\linewidth]{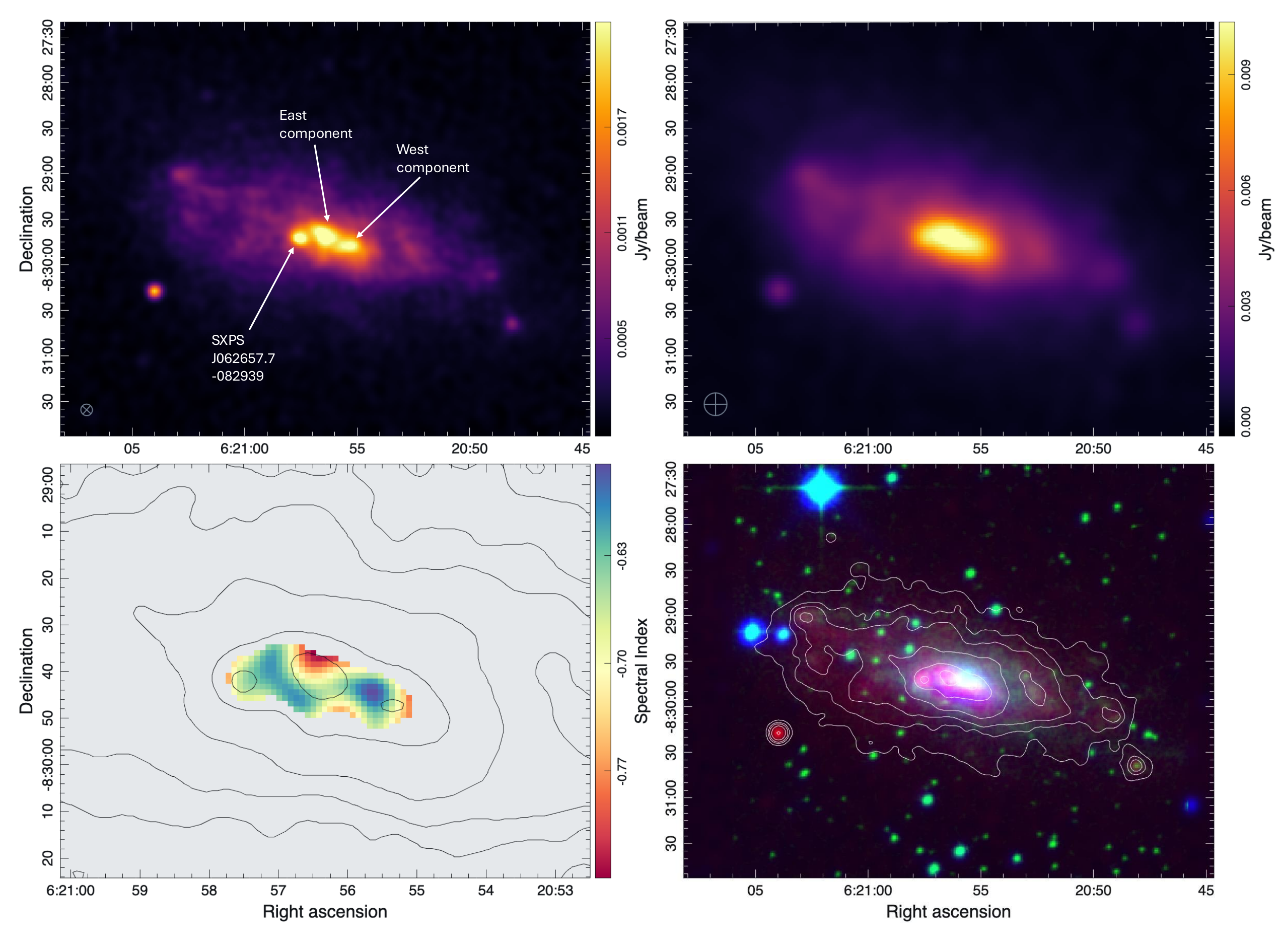}
    \caption{Multi-frequency and spectral index maps of the galaxy \ugca\ (\ugcb). {\bf Top left:} MeerKAT 1.3\,GHz radio continuum image. The radio components to the east and west of the galaxy core and the X-ray source 2SXPS~J062057.7--082939 are annotated. The synthesized beam (7\farcs68$\times$7\farcs52, P.A.\,=\,--40\arcdeg) is shown in the bottom left corner. The image has a local \ac{RMS} noise level of $\sim$15$\,\mu$Jy\,beam$^{-1}$. {\bf Top right:} \ac{ASKAP} 944\,MHz radio continuum image. The 15\arcsec\ synthesized beam is shown in the bottom left corner. The image has a local \ac{RMS} noise level of $\sim$50\,$\mu$Jy beam$^{-1}$. {\bf Bottom left:} Spectral index map zoomed to show only the central emission. The map was generated using the high-resolution \ac{ASKAP} 944\,MHz image, the MeerKAT 1.3\,GHz image, and the 3-epoch averaged \ac{VLASS} 3~GHz image as described in Sec.~\ref{subsubsec:ugc_Discussion}. The final image was convolved to 10\arcsec. {\bf Bottom right:} Multi-frequency RGB image, consisting of MeerKAT 1.3\,GHz (red), optical DSS2 $R$-band (green), and WISE 3.4\,$\mu$m (blue) images. All images are linearly scaled, and the bottom row includes MeerKAT 1.3\,GHz radio contours at levels of 10, 20, 30, 50, 100, and 150\,$\sigma$, where 1\,$\sigma$\,=\,15\,$\mu$Jy\,beam$^{-1}$.}
    \label{fig:fig2}
\end{figure*}

\section{Results and Discussion}
 \label{sec:resultsanddiscussion}

\subsection{Catalog generation}
\label{subsec:catalog_Generation}

We use the source finding software \textsc{AEGEAN} \citep{hancock2018} to catalog \ac{ASKAP}-\ac{EMU} radio sources above 10$\sigma$ within the 68\% certainty search area. The available \ac{EMU} data only covered approximately half of the 99\% certainty region, and so the 68\% certainty region was used. Note that part of the north-western region is not covered (Figure~\ref{fig:fig1}). We found \totnumps\ radio point sources, and seven extended sources at 15\arcsec\ resolution, which we manually measured by fitting regions using the CARTA imaging software, following the methodology described in~\citet{2022MNRAS.512..265F, 2024PASA...41..112F, 2024MNRAS.534.2918S, 2025PASA...42...32B}. We estimate a 10\% error for all \ac{ASKAP} flux density measurements (Hopkins et al., submitted.).

We use the \ac{EMU} catalog as a base to cross-match with other catalogs using a 2\arcsec\ matching radius. We calculate the spectral indices $\alpha$, defined as $S\propto\nu^{\alpha}$ \citep[]{book1}, for all sources with more than two flux density measurements, and cross-match with redshift surveys to find distances. We calculate spectral indices for 185 sources and find cataloged redshifts for one point source \citep{2009MNRAS.399..683J} and three extended sources \citep{Koribalski2004, 2009MNRAS.399..683J, 2014ApJS..210....9B}. A table of the documented information is provided in Appendix~\ref{Appendix:Sample}.

\subsection{Objects of interest}
\label{subsec:OoI}

We find several objects of interest. The potential blazar counterpart, PMN\,J0606--0724, analyzed by~\cite{2025arXiv250208484K}, is not discussed here as it is located outside of the \ac{EMU} tile coverage.

\subsubsection{Radio spectra candidates}
\label{subsubsec:OSI_Discussion}

We calculated spectral indices for 185 point sources with a mean value of $\alpha$\,=\,$-$0.75$\pm$0.12 (SD\,=\,0.39$\pm$0.15). This value and distribution match the observed average for extragalactic radio point sources ~\citep{2021MNRAS.506.3540P, 2021MNRAS.507.2885F}, indicating that the sources have predominantly extragalactic origin. 

Blazars are one of the expected astrophysical high-energy neutrino sources, and are typically radio-continuum variable with flat spectral indices~\citep{book1}. We find ten blazar candidates based on their spectra, seven that exhibit a \ac{GPS}-like spectrum~\citep{2018MNRAS.477..578C} (EMU~J061335--080331, EMU~J061750--083920, EMU~J061831--075415, EMU~J061835--081043, EMU~J061848--074327, EMU~J062215--075522, and EMU~J062245--072330 (see Sec.~\ref{subsubsec:VLASS})), and three that exhibit possible spectral variability (EMU~J061227--080810, EMU~J061415--083920, and EMU~J062244--072334). These blazar candidates are also possible producers of the \km\ event.

\subsubsection{\ugcb\ (\ugca; EMU~J062056$-$082949)}
\label{subsubsec:ugc_Discussion}

The most prominent radio source in the field is the nearby Scd type galaxy \ugca\ (HIPASS~J0620--08, see Appendix~\ref{App2:HIfig}), with a systemic velocity of $v_{\rm sys} = 732 \pm 2$ km\,s$^{-1}$ \citep{Koribalski2004} and a distance of $D=8.3\pm0.6$\,Mpc. We give it the nickname \ugcb.

We measure a radio-continuum extent of 210\arcsec$\times$90\arcsec\ (8.4$\times$3.6\,kpc) in the \ac{ASKAP} 944\,MHz image (Figure~\ref{fig:fig2}, top-right) and detect two resolved hotspot-like features in the 1.3\,GHz MeerKAT image. These indicate potential \ac{AGN} jets oriented within the galaxy plane (marked East and West in Figure~\ref{fig:fig2}, top-left). We detect linear polarization in the \ac{POSSUM} data. Unfortunately, this area is contaminated with bright, smooth Galactic foreground emission at overlapping \ac{RM}s, meaning we cannot easily separate the features, and so this polarization data is not used in this analysis. We combine our cataloged \ac{ASKAP} flux densities (Appendix~\ref{Appendix:Sample}) with \ac{PMN}~\citep[$S_{4850\,\text{MHz}}$\,=\,0.084$\pm$0.011\,Jy] {1995ApJS...97..347G} and MeerKAT~\citep[$S_{1280\,\text{MHz}}$\,=\,0.201$\pm$0.020\,Jy; B.S.\,=\,7.68\arcsec$\times$7.52$\arcsec$]{Condon2021}, and calculate an $\alpha\,=\,-$0.61$\pm$0.01 and a radio luminosity $L_{1\,\text{GHz}}$\,=\,1.9$\times$10$^{21}$\,W\,Hz$^{-1}$. This indicates non-thermal emission, consistent with the presence of possible \ac{CR} accelerators within \ugcb.

We generate a spectral index map with high-resolution \ac{EMU}, MeerKAT, and 3-epoch averaged \ac{VLASS} data using the \textsc{miriad}~\citep{Sault1995} \textsc{maths} function. We convolve all images to a common beam size of 10\arcsec$\times$10\arcsec\ and calculate the spectral index per pixel. The \ac{VLASS} image is only reliable for the inner emission due to missing short spacing, so the images were cut at a 2\,mJy\,beam$^{-1}$ level. We measure an average $\alpha=-0.67\pm0.07$ over the entire region shown (Figure~\ref{fig:fig2}, middle left), with the west component being the flattest $\alpha=-0.57\pm0.06$. This is consistent with a central engine of higher energy injection and particle acceleration in the west component.

We calculate \ugcb's IR colors using AllWISE data~\citep{2013wise.rept....1C} as W1$-$W2\,=\,0.226 and W2$-$W3\,=\,3.839. Comparing with the color-color diagram of~\citet[their Figure~12]{2010AJ....140.1868W}, we find \ugcb\ located in the region common to spiral galaxies, starburst galaxies, and \acp{LIRG}. Being in the starburst region indicates that \ugcb\ may be experiencing a starburst, potentially caused by \ac{AGN} feedback, which provides another potential site of particle acceleration \citep{2024APh...16202990A,2021JInst..16C2016M}. While we lack direct observational evidence for star-forming galaxies as \ac{UHE} neutrino sources~\citep{2017ApJ...836...47B}, there are numerous models which predict these areas can produce neutrinos up to PeV energies~\citep{2015MNRAS.453..113B}.

We detect an offset between the central radio hotspots, IR and optical images, suggesting possible inner knots of a radio jet rather than the galaxy nucleus \citep{2022MNRAS.516.1865V}. If the emission is caused by jets, it may indicate areas of higher energy if the jets are oriented within the galaxy plane and thus pushing through denser material. It also suggests that the emission may instead be similar to a superbubble,~\citep[e.g. 30-Doradus~C in the Large Magellanic Cloud;][]{2015A&A...573A..73K,2017ApJ...843...61S,2019A&A...621A.138K,2021ApJ...918...36Y}, microquasars \citep[e.g. Galactic SS~433 or NGC\,7793's S26;][]{2010MNRAS.409..541S, 2024Sci...383..402H}, recent core-collapse \ac{SN} event(s) \citep{2022EPJC...82..317A,2023PJAB...99..460A}, or massive star clusters \citep[e.g. Westerlund~1;][]{2025arXiv250112990H}. All of these are areas of potential particle acceleration and \ac{CR} or neutrino origin. The likely presence of dense material within the central galaxy provides a natural target for neutrino production by the \acp{CR}.

We also suggest the \ac{XRB} 2SXPS~J062057.7--082939~\citep[see Figure~\ref{fig:fig1} top-left;][]{2021PASJ...73.1315I,2022MNRAS.509.1587W} located $\sim$25\arcsec\ from \ugcb's center as a possible origin for \km. This source is detected in 2SXP with an X-ray luminosity $L\,=\,8.71\times10^{38}$\,erg\,s$^{-1}$ (2--10 keV range), but is not detected in eRASS1 \citep{2024A&A...682A..34M}, indicating that it is likely highly X-ray variable, as expected from an active \ac{XRB}. We see a possible radio counterpart in the MeerKAT image and estimate a radio luminosity $L_{1\,\text{GHz}}\sim$10$^{19}$\,W\,Hz$^{-1}$. This would make the object superluminous for an \ac{XRB} radio counterpart, indicating it may be an area of \ac{UHE} particle acceleration~\citep{2022MNRAS.513.2708R}. Using the \ac{ASKAP}, MeerKAT and \ac{VLASS} 3-epoch combined image, we estimate a spectral index, $\alpha\,=\,-0.5\pm0.2$, and detect no observable variability between the three \ac{VLASS} epochs. We also note three \ac{XRB} candidates in \ugcb's outer field (2SXPS~J062053.1--083026, 1eRASS~J062054.9--083006, 1eRASS~J062047.1--083026) that have no corresponding radio emission.

\ugcb\ is unlikely to contain a blazar due to its steeper spectral index and lack of detected variability. However, it hosts several potential areas of \ac{UHE} particle acceleration. Due to particle escape as in the commonly used ``leaky box'' models, the spectrum of the diffuse \acp{CR} within spiral galaxies is likely different than the \ac{CR} spectra injected from the accelerators. Furthermore, reacceleration of \acp{CR}, such as for example in the so-called ``espresso'' model, where the \acp{CR} produced at other locations of the galaxy are reaccelerated within the jets of a central \ac{AGN} engine, can also change the resulting \ac{CR} distribution and contribute to the production of \ac{UHE}\acp{CR} \citep{2015ApJ...811L..38C}. These processes might be identified by differentiating areas of different spectral indices within a galaxy, such as done by \citet{1995ApJ...445..173D, 2000immm.proc..179D}. However, our radio data is unfortunately not of sufficiently high resolution for such a detailed spectral analysis. The proposed \ac{AGN} jets are a potential origin, but they appear oriented away from us, and the \ac{SFR} is inconclusive. These jet orientations have been observed in other galaxies, some of which are possible high-energy neutrino sources, e.g. NGC\,1068~\citep{2024NatAs...8.1077P}, NGC\,7469~\citep{2024arXiv240303752S}, and NGC\,5938~(Zakir~et~al.~in~prep). It is possible that these jets are emitting \ac{UHE}-\acp{CR} into surrounding space; by interacting with ambient gas or photons, they can produce neutrinos. This raises the possibility that the neutrino direction may not point directly to the particle acceleration site.

To assess the probability of chance association of \ugcb, we used the bright galaxy number counts \citep{2001AJ....122.1104Y}, which implies 0.4 galaxies with a similar or brighter magnitude in a 1\fdg5 radius. This can mean a likely chance association, but we note that galaxies with radio properties similar to that of \ugcb\ are rarer than the total number counts.

\subsubsection{\wisegalb\ (\wisegala)}
 \label{subsubsec:wisegal_Discussion}

\begin{figure}
     \centering
     \includegraphics[width=0.51\linewidth]{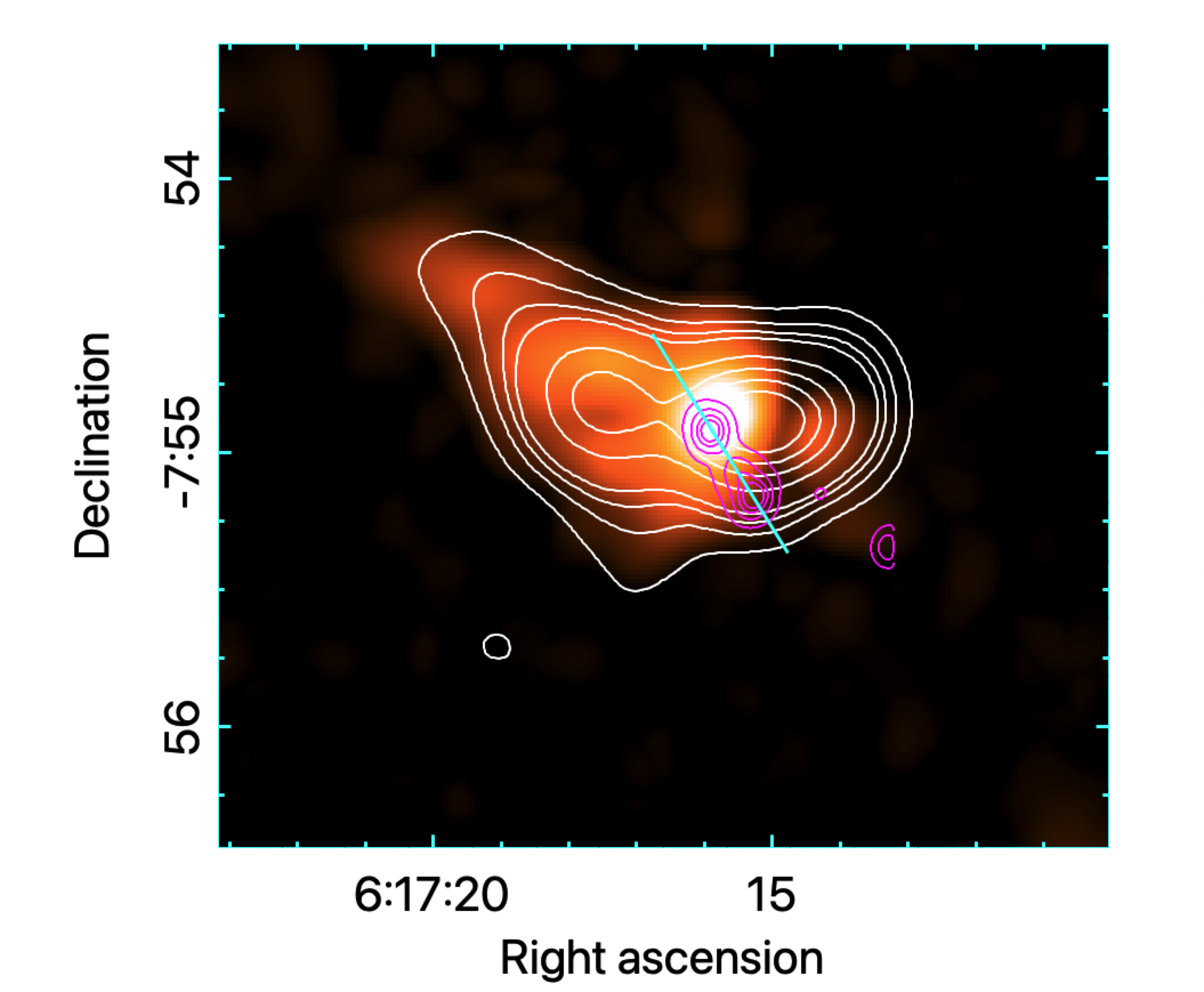}
     \includegraphics[width=0.43\linewidth]{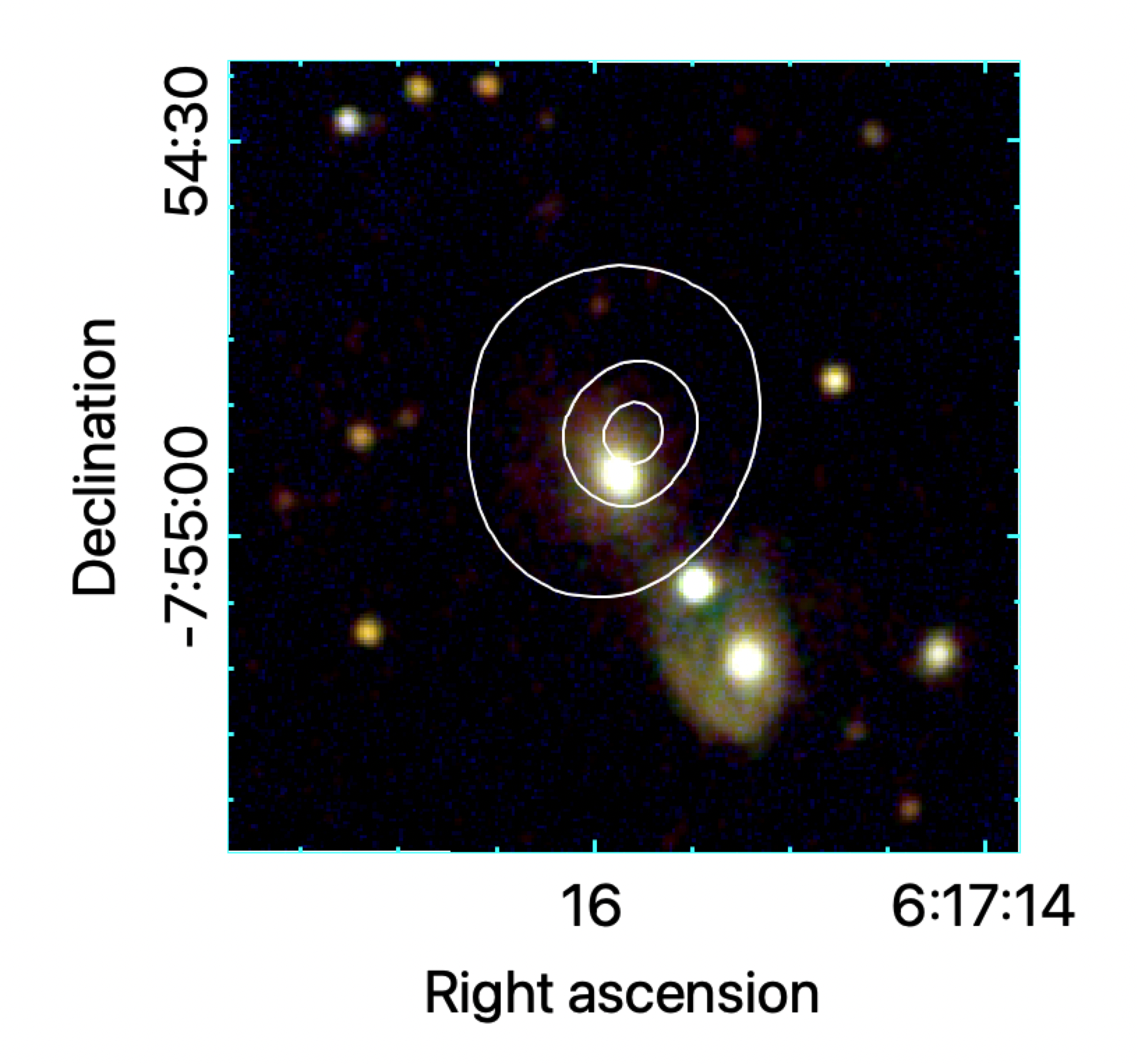}
     \caption{Left: polarized intensity of \wisegalb\ at 944~MHz, overlaid with white contours of total intensity at (0.5, 1, 1.5, 2, 4,...)\,mJy\,beam$^{-1}$.  The cyan line indicates the orientation of the inferred magnetic field at the polarized peak after Faraday rotation correction, and the magenta contours are of the WISE~W1 image. Right: PanSTARRS image, combining the $z, i,$ and $r$ frames, overlaid with contours of polarized intensity at 0.5, 0.8 and 0.9\,mJy\,beam$^{-1}$. The faint optical envelope is seen, in more burned-out images, to encompass all three galaxies.}
     \label{fig:hebe}
\end{figure}

The closest extended radio source to the target position is found $\sim$7\arcmin\ to its south. Called \wisegalb\ here, it is a double radio galaxy with diffuse lobes and an extraordinary slightly resolved polarized component, offset by 3\arcsec\ along the direction of its major axis from the IR galaxy \wisegala\ (see Figure \ref{fig:hebe}). At a redshift of $z$\,=\,0.125$\pm$0.017 \citep[$D$\,=\,553.89$\pm$38.77\,pc; ][]{2014ApJS..210....9B}, this offset corresponds to 8\,kpc. No compact component is visible in the total intensity image, but we made a very rough estimate of its contribution by forcing a fit to a Gaussian component at that position, along with additional components for the lobes. This yielded a flux of $\approx$1.5\,mJy/(18\arcsec\,beam).

The bright peak polarized intensity is 0.8\,mJy\,beam$^{-1}$ above its local background. There is clear depolarization, consistent, e.g., with a Burn-slab model with an rms scatter of $\sim5\,\frac{\rm rad}{\rm m^2}$. After correction for depolarization, this corresponds to an apparent remarkable $\ge$100\% polarization, varying only slightly with the unknown spectral index. Even core-dominated quasars have typical fractional polarizations of only a few percent \citep{2023MNRAS.520.6053P}, although the resolved components of pc-scale \ac{VLBI} jets can reach levels of 10s of percent \citep[e.g.,][]{2005AJ....130.1418J}. The rotation measure of the polarized peak is $59\pm1\,\frac{\rm rad}{\rm m^2}$, close to the value of the local Galactic foreground, \citep[$\approx52\,\frac{\rm rad}{\rm m^2}$;][]{2022A&A...657A..43H}, so there is no evidence for a large, additional Faraday component local to \wisegalb. 

At 8\,kpc from the presumed \ac{AGN} core in \wisegala, \wisegalb's central polarized component could also be a jet, which is only slightly resolved here, although its direction would be almost perpendicular to the more extended emission. 
We estimate an integrated spectral index $\alpha\,=\,-0.65\pm0.02$ for \wisegalb\ (see Table~\ref{table:sample_cat}), although we are not able to measure any spatial-spectral structure, especially of the polarized core. 

There were four \ac{ASKAP} observations of the field spanning December~11$^{\rm th}$, 2023 through April~13$^{\rm th}$, 2024, three of which were rejected because of instrumental problems.  Nonetheless, we looked for variability in the peak polarized intensity and found it to be $<10\%$. In total intensity, the quality of the data prevented any useful limits.

\wisegala\ is the northernmost member of a triplet of galaxies, which can be seen in PanSTARRS to be embedded in a common faint envelope.  The envelope's southern boundary has a curious bow shape, perhaps indicating dynamical interactions within this system.  

It is unclear whether \wisegalb\ is related to the class of blazars or other core-dominated radio \ac{AGN}. Blazars are of interest as possible sources of cosmic neutrinos \citep{2024ApJ...964....3A}, so further investigation of \wisegalb\ is warranted. 
One interesting model to explore for \wisegalb\ would be the ``espresso'' scheme \citep{2015ApJ...811L..38C} using relativistic jets for the generation of \ac{UHE}\acp{CR}. Monitoring in radio polarization and searching for possible optical polarization in \wisegala\ would be especially valuable.

\subsubsection{VLASS investigation and Narcissus (EMU~J062248$-$072246) }
 \label{subsubsec:VLASS}

We search for \ac{VLASS} counterparts for all cataloged \ac{ASKAP} sources. We then average the images from the three \ac{VLASS} epochs and convolve them to the \ac{ASKAP} resolution of 15\arcsec. We record the peak flux densities as \ac{VLASS} is insensitive to extended emission. We measure the spectral index between \ac{ASKAP} and \ac{VLASS} and find $\alpha\sim-1$ for the brightest \ac{ASKAP} sources, slightly steeper than expected~\citep{2021MNRAS.506.3540P, 2024MNRAS.529.2443C} likely due to missing \ac{VLASS} flux densities, and flatter and inverted spectral indices for dimmer \ac{ASKAP} sources ($<$ 5\,mJy) due to our 3$\sigma$ \ac{VLASS} cutoff. We find six sources with $S_{944}>5$~mJy and $\alpha > 0$ (inverted), marked with asterisks in Appendix~\ref{Appendix:Sample}, which deserve further scrutiny.

We examine the sources with the largest variability in the three \ac{VLASS} epochs. The median differences between the full-resolution epoch pairs for this \ac{VLASS} tile can be modeled as a noise contribution of $\sim$0.2~mJy\,beam$^{-1}$, and a flux normalization error of $\sim$5\%, arising from instrumental effects including incomplete cleaning. We visually examine 20 sources showing the highest fractional variability integrating the flux density in a 20\arcsec\ box. We investigate a number of sources which show significant variability.

\begin{figure*}
    \centering
    \includegraphics[width=1\linewidth]{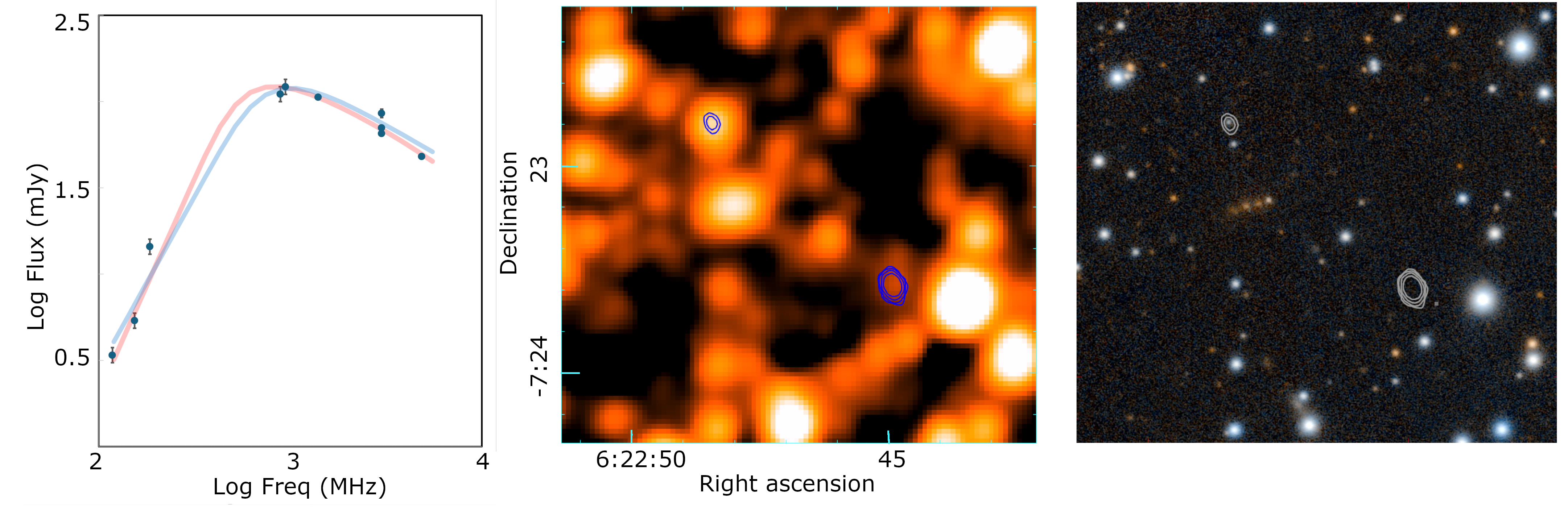}
    \caption{ Left: Broadband spectrum of \namethree\ (EMU~J062244$-$072333, including minor contributions from EMU~J062248$-$072246 low frequencies, from the larger beams). Values from all three \ac{VLASS} epochs are shown. The illustrative spectra are for a homogeneous self-absorbed synchrotron source, with an optical depth of 1 at 600\,MHz and an optically thin index of --0.5 (pink), and synchrotron emission from a sub-relativistic shock of {\it thermal} electrons at frequencies below the thermal cooling threshold, with an optically thin slope of --0.15 (blue). Middle: Overlay of the two radio sources from the \ac{VLASS} map averaged over all epochs on the WISE~W1 image. Right: Overlay on the multi-band PanSTARRS image. 
    }
    \label{fig:larry}
\end{figure*}

The most obviously variable source across the \ac{VLASS} epochs (2017~November~30, 2020~October~9, and 2023~March~10) is the southern source in Figure~\ref{fig:larry}, (EMU~J062248$-$072246, hereinafter \namethree). At the native \ac{VLASS} 3\arcsec\ resolution, its peak flux densities are 85.4, 70.4, and 65.2~mJy\,beam$^{-1}$ with errors of 0.15~mJy\,beam$^{-1}$. The northern source (EMU~J062248$-$072246) also appears variable, with peak flux densities of 2.1, 3.7 and 1.3~mJy\,beam$^{-1}$. Both components are consistent with being unresolved within the errors. We also search within the \ac{EMU} data, but detect no variability for \namethree\ during the 5-hr observation. Both sources have WISE counterparts (WISEA~J062244.90--072334.9 -- south, and WISEA~J062248.41--072247.9 -- north), with colors consistent with \acp{AGN}, as expected for radio-selected samples \citep[e.g.,][]{2025MNRAS.536.3488W}. To better separate the adjacent bright WISE source in the south, we looked at the unWISE colors, which yield W1--W2=1.3, consistent with an \ac{AGN}. Approximately 4\% of compact sources at 3~GHz exhibit variability at the $\sim$30\% level \citep{2016ApJ...818..105M}, and are expected to be associated with \ac{AGN}, so these results themselves are not unusual.

What is unusual is the broadband spectra of \namethree, when we combine 4.85\,GHz data \citep{1995ApJS...97..347G} with our catalog data (Figure~\ref{fig:larry}, left). We show two possible models. The first model, with the maximum possible low-frequency slope of $\alpha=+2.5$, corresponds to a compact, homogeneous source with synchrotron self-absorption. At the high, optically thin part of the spectrum, where the variability is observed, the slope is $\alpha$=--0.5, consistent with the limit for strong (non-relativistic) Fermi~I shock acceleration \citep{book1}. Although an \ac{AGN} connection for \namethree\ is attractive, we cannot rule out a galactic origin. We therefore show an alternative model in Figure~\ref{fig:larry}, with a low-frequency slope of $\alpha=+2$. This is similar to shock models explored by \cite{MQ24}, e.g., their Fig.~3(b), used to describe explosive events such as SN1998bw and AT2018cow. Thermal absorption is another possibility for the extremely low-frequency slope.

Independent of the origins of \namethree, Galactic or extragalactic, the steep low-frequency slope indicates that the emission is dominated by a compact, homogeneous component. This is built-in to explosive models relying on a single shock, whereas inhomogeneities, or multiple components, would broaden the spectrum and flatten the slope. However, it is not clear how the need for a compact, homogeneous source applies to \namethree since the broadband spectra span many years of observations. Further multi-frequency monitoring and \ac{VLBI} observations are necessary to separate the spectra of the short- and long-term components and to compare, e.g., to the evolutionary scenarios, e.g., of \cite{2021ApJ...923L..14M}.

The third \ac{VLASS} epoch was observed less than one month after the \km\ event, so we conduct an additional search for variability between the \ac{VLASS} epochs, independent of the \ac{ASKAP} base catalog, based on an approach similar to \citet{2019MNRAS.490.4898H} and \citet{2021MNRAS.501.6139R}. We identify two variable sources with $V_{S} = \Delta S/\sigma_{\Delta S} > 4\sigma (V_{S, \text{population}})$ and $m = \Delta S/\overline{S} > 0.26 $. The first source is the rotating variable radio star 1RXS~J061542.9$-$072339\footnote{\url{https://radiostars.org/}} \citep{2024PASA...41...84D} which brightens from flux densities of $2.4 \pm 0.3\,$ and $2.1 \pm 0.3\,$mJy in the first two epochs, to $4.9 \pm 0.6\,$mJy in the third epoch. The second source is the \ac{AGN}, WISEA~J062248.41$-$072247.9, which has flux densities of 2.1$\pm$0.3, 3.7$\pm$0.4 and $1.4\pm$0.2\,mJy, in the three \ac{VLASS} epochs, respectively. Its radio variability and the quasar-like IR colors are consistent with blazars, which are known neutrino emitters \citep{2018Sci...361.1378I}.

\subsubsection{VAST investigation}
 \label{subsubsec:VAST}

The \km\ event is covered by the \ac{ASKAP} Variables and Slow Transients (VAST) extragalactic survey \citep{2021PASA...38...54M}. The survey began in June~2023 and covers $\sim 10,000$\,deg$^2$ at a central frequency of 888\,MHz to a noise level of $250\,\mu$Jy approximately every two months.

We queried the VAST transient detection pipeline \citep{adam_stewart_2024_14048598} for sources within the localization region using {\sc vast-tools} \citep{adam_stewart_2024_13131754}. We applied standard quality cuts to remove artifacts, requiring the nearest neighbor to be separated by $>15\arcsec$; a peak-integrated flux ratio of $<1.4$ and at least one detection with an S/N greater than seven. 44 sources satisfy these criteria, but after manually inspecting all light curves, none exhibit sufficient variability to be associated with the neutrino event.

Separately, we have searched for variability from the two \ac{EMU} sources (shown in Figure~\ref{fig:larry}) of interest reported in this work. EMU~J062248--072246 is not detected in our survey, which is not surprising given the relative sensitivity. EMU~J062244--072333 is detected in all 10 observations which cover it, with a typical integrated flux density of $\sim 110\,$mJy. It exhibits minimal variability, with a modulation index of 8\% across the observation, and no obvious trends, with any intrinsic variability inseparable from stochastic variability induced by refractive interstellar scintillation.

\section{Conclusion}
 \label{sec:Conclusion}

We catalog \ac{ASKAP} and \ac{VLASS} radio sources in the 68\% confidence region of the \ac{UHE} neutrino event, \km. 
We identify ten possible blazar candidates in the field that could produce \ac{UHE} neutrinos. We further analyze three individual objects, the nearby spiral galaxy UGCA~127 (Phaedra), the radio galaxy (\wisegalb) closest to the geometric search area center and an unusual compact radio source (Narcissus). Phaedra is the most prominent radio source in the field and is a star-forming galaxy. However, jets may be interacting with the galaxy plane itself, causing starburst regions. This possibility, as well as that of \acp{SNR} and microquasars within the galaxy, and the well-known \ac{XRB}, provide areas of possible neutrino generation within Phaedra. \wisegalb~ contains an unusual, very highly polarized compact component, likely associated with one of a triplet of galaxies in a common envelope. Narcissus is a radio source that displays radio variability in \ac{VLASS}, originating in a compact, heavily absorbed \ac{AGN} or a more local explosive event. These properties are consistent with known neutrino produces, making Narcissus a plausible candidate for the neutrino origin. Both Hebe and Narcissus are clearly associated with an explosive activity that could produce \ac{UHE} neutrinos, making them the most likely candidates. Phaedra shows evidence of more high-energy emission than other galaxies, making it a possible candidate, albeit this evidence is somewhat weaker than for the others. All of these systems deserve further investigation as possible origins for \km.

\section{Acknowledgments} 
 \label{sec:Acknowledgements}

This scientific work uses data obtained from Inyarrimanha Ilgari Bundara / the Murchison Radio-astronomy Observatory. We acknowledge the Wajarri Yamaji People as the Traditional Owners and native title holders of the Observatory site. The \ac{CSIRO}’s \ac{ASKAP} radio telescope is part of the Australia Telescope National Facility\footnote{\url{https://ror.org/05qajvd42}}. Operation of \ac{ASKAP} is funded by the Australian Government with support from the National Collaborative Research Infrastructure Strategy. \ac{ASKAP} uses the resources of the Pawsey Supercomputing Research Centre. Establishment of \ac{ASKAP}, Inyarrimanha Ilgari Bundara, the \ac{CSIRO} Murchison Radio-astronomy Observatory and the Pawsey Supercomputing Research Centre are initiatives of the Australian Government, with support from the Government of Western Australia and the Science and Industry Endowment Fund. 
The MeerKAT telescope is operated by the South African Radio Astronomy Observatory, which is a facility of the National Research Foundation, an agency of the Department of Science and Innovation.
We thank C.W. James for discussion relating to the origin of the neutrino event and Yuri Y. Kovalev for the radio-continuum discussion.
This research was funded in part by the National Science Centre, Poland, grant Nos.2018/30/E/ST9/00208 and 2023/49/B/ST9/00066. 
This research was supported by the Sydney Informatics Hub (SIH), a core research facility at the University of Sydney. This work was also supported by software support resources awarded under the Astronomy Data and Computing Services (ADACS) Merit Allocation Program. ADACS is funded from the Astronomy National Collaborative Research Infrastructure Strategy (NCRIS) allocation provided by the Australian Government and managed by Astronomy Australia Limited (AAL).
We thank an anonymous referee for excellent comments and suggestions, which greatly improved this paper.

\section{Data Availability} 
\label{sec:Data Availability}

The \ac{ASKAP}-\ac{EMU} data used in this work are available through the \ac{CASDA} (\url{https://data.csiro.au/domain/casdaObservation}). The full table of \ac{ASKAP} \ac{EMU} sources generated for this paper will be made available on VizieR (\url{https://vizier.cds.unistra.fr/viz-bin/VizieR}) upon submission.

\bibliography{KM3}

\begin{thebibliography}{}
\expandafter\ifx\csname natexlab\endcsname\relax\def\natexlab#1{#1}\fi
\providecommand{\url}[1]{\href{#1}{#1}}
\providecommand{\dodoi}[1]{doi:~\href{http://doi.org/#1}{\nolinkurl{#1}}}
\providecommand{\doeprint}[1]{\href{http://ascl.net/#1}{\nolinkurl{http://ascl.net/#1}}}
\providecommand{\doarXiv}[1]{\href{https://arxiv.org/abs/#1}{\nolinkurl{https://arxiv.org/abs/#1}}}

\bibitem[{{Adriani} {et~al.}(2025){Adriani}, {Aiello}, {Albert}, {Alhebsi},
  {Alshamsi}, {Alves Garre}, {Ambrosone}, {Ameli}, {Andre}, {Aphecetche},
  {Ardid}, {Ardid}, {Aublin}, {Badaracco}, {Bailly-Salins},
  {Barda{\v{c}}ov{\'a}}, {Baret}, {Bariego-Quintana}, {Becherini}, {Bendahman},
  {Benfenati Gualandi}, {Benhassi}, {Bennani}, {Benoit}, {Berbee}, {Berti},
  {Bertin}, {Betti}, {Biagi}, {Boettcher}, {Bonanno}, {Bottai}, {Bouasla},
  {Boumaaza}, {Bouta}, {Bouwhuis}, {Bozza}, {Bozza}, {Br{\^a}nzas},
  {Bretaudeau}, {Breuhaus}, {Bruijn}, {Brunner}, {Bruno}, {Buis}, {Buompane},
  {Busto}, {Caiffi}, {Calvo}, {Capone}, {Carenini}, {Carretero}, {Cartraud},
  {Castaldi}, {Cecchini}, {Celli}, {Cerisy}, {Chabab}, {Chen}, {Cherubini},
  {Chiarusi}, {Circella}, {Clark}, {Cocimano}, {Coelho}, {Coleiro},
  {Condorelli}, {Coniglione}, {Coyle}, {Creusot}, {Cuttone}, {Dallier}, {De
  Benedittis}, {De Wasseige}, {Decoene}, {Deguire}, {Del Rosso}, {Di Mauro},
  {Di Palma}, {D{\'\i}az}, {Diego-Tortosa}, {Distefano}, {Domi}, {Donzaud},
  {Dornic}, {Drakopoulou}, {Drouhin}, {Ducoin}, {Duverne}, {Dvornick{\'y}},
  {Eberl}, {Eckerov{\'a}}, {Eddymaoui}, {van Eeden}, {Eff}, {van Eijk}, {El
  Bojaddaini}, {El Hedri}, {El Mentawi}, {Ellajosyula}, {Enzenh{\"o}fer},
  {Ferrara}, {Filipovi{\'c}}, {Filippini}, {Franciotti}, {Fusco}, {Gal},
  {Garc}, {Garcia Soto}, {Gatius Oliver}, {Gei}, {Genton}, {Ghaddari},
  {Gialanella}, {Gibson}, {Giorgio}, {Goos}, {Goswami}, {Gozzini}, {Gracia},
  {Guidi}, {Guillon}, {Guti}, {Haack}, {van Haren}, {Heijboer}, {Hennig},
  {Hern}, {Idrissi}, {Idrissi Ibnsalih}, {Illuminati}, {Janik}, {Joly}, {de
  Jong}, {de Jong}, {Jung}, {Kalaczy{\'n}ski}, {Keegans}, {Kikvadze},
  {Kistauri}, {Kopper}, {Kouchner}, {Kovalev}, {Krupa}, {Kueviakoe},
  {Kulikovskiy}, {Kvatadze}, {Labalme}, {Lahmann}, {Lamoureux}, {Larosa},
  {Lastoria}, {Lazar}, {Lazo}, {Le Stum}, {Lehaut}, {Lema}, {Leonora},
  {Lessing}, {Levi}, {Clark}, {Longhitano}, {Magnani}, {Majumdar}, {Malerba},
  {Mamedov}, {Manfreda}, {Manousakis}, {Marconi}, {Margiotta}, {Marinelli},
  {Markou}, {Martin}, {Mastrodicasa}, {Mastroianni}, {Mauro}, {Mehta},
  {Meskar}, {Miele}, {Migliozzi}, {Migneco}, {Mitsou}, {Mollo},
  {Morales-Gallegos}, {Mori}, {Moussa}, {Mozun Mateo}, {Muller}, {Musone},
  {Musumeci}, {Navas}, {Nayerhoda}, {Nicolau}, {Nkosi}, {/'O. Fearraigh},
  {Oliviero}, {Orlando}, {Oukacha}, {Pacini}, \&
  {Paesani}}]{2025arXiv250208387A}
{Adriani}, O., {Aiello}, S., {Albert}, A., {et~al.} 2025, arXiv e-prints,
  arXiv:2502.08387.
\newblock \doarXiv{2502.08387}

\bibitem[{{Aiello} {et~al.}(2022){Aiello}, {Albert}, {Alshamsi}, {Alves Garre},
  {Aly}, {Ambrosone}, {Ameli}, {Andre}, {Androulakis}, {Anghinolfi}, {Anguita},
  {Ardid}, {Ardid}, {Aublin}, {Bagatelas}, {Baret}, {Pree}, {Bendahman},
  {Benfenati}, {Berbee}, {Berg}, {Bertin}, {Biagi}, {Boettcher}, {Cabo},
  {Boumaaza}, {Bouta}, {Bouwhuis}, {Bozza}, {Br{\^a}nza{\c{s}}}, {Bruijn},
  {Brunner}, {Bruno}, {Buis}, {Buompane}, {Busto}, {Caiffi}, {Calvo},
  {Campion}, {Capone}, {Carretero}, {Castaldi}, {Celli}, {Chabab}, {Chau},
  {Chen}, {Cherubini}, {Chiarella}, {Chiarusi}, {Circella}, {Cocimano},
  {Coelho}, {Coleiro}, {Molla}, {Coniglione}, {Coyle}, {Creusot}, {Cruz},
  {Cuttone}, {Dallier}, {De Martino}, {Palma}, {D{\'\i}az}, {Diego-Tortosa},
  {Distefano}, {Domi}, {Donzaud}, {Dornic}, {D{\"o}rr}, {Drouhin}, {Eberl},
  {Eddyamoui}, {Eeden}, {Eijk}, {Bojaddaini}, {Hedri}, {Enzenh{\"o}fer},
  {Espinosa}, {Fermani}, {Ferrara}, {Filipovi{\'c}}, {Filippini}, {Fusco},
  {Gal}, {M{\'e}ndez}, {Soto}, {Garufi}, {Gatelet}, {Oliver},
  {Gei{\ss}elbrecht}, {Gialanella}, {Giorgio}, {Gozzini}, {Gracia}, {Graf},
  {Grella}, {Guderian}, {Guidi}, {Guillon}, {Guti{\'e}rrez}, {Haefner},
  {Hallmann}, {Hamdaoui}, {Haren}, {Heijboer}, {Hekalo}, {Hennig},
  {Hern{\'a}ndez-Rey}, {Hofest{\"a}dt}, {Huang}, {Ibnsalih}, {Illuminati},
  {James}, {Janezashvili}, {de Jong}, {de Jong}, {Jung}, {Kalaczy{\'n}ski},
  {Kalekin}, {Katz}, {Chowdhury}, {Kistauri}, {Knaap}, {Kooijman}, {Kouchner},
  {Kulikovskiy}, {Labalme}, {Lahmann}, {Lamoureux}, {Larosa}, {Lastoria},
  {Lazo}, {Breton}, {Stum}, {Lehaut}, {Leonardi}, {Leone}, {Leonora},
  {Lessing}, {Levi}, {Lincetto}, {Clark}, {Lipreau}, {Alvarez}, {Longhitano},
  {Lopez-Coto}, {Maderer}, {Majumdar}, {Ma{\'n}czak}, {Margiotta}, {Marinelli},
  {Markou}, {Martin}, {Mart{\'\i}nez-Mora}, {Martini}, {Marzaioli},
  {Mastroianni}, {Melis}, {Miele}, {Migliozzi}, {Migneco}, {Mijakowski},
  {Miranda}, {Mollo}, {Moser}, {Moussa}, {Muller}, {Musumeci}, {Nauta},
  {Navas}, {Nicolau}, {Nkosi}, {Fearraigh}, {O'Sullivan}, {Organokov},
  {Orlando}, {Gonz{\'a}lez}, {Papalashvili}, {Papaleo}, {P{\u{a}}un},
  {P{\u{a}}v{\u{a}}la{\c{s}}}, {Pellegrino}, {Perrin-Terrin}, {Pestel},
  {Piattelli}, {Pieterse}, {Pisanti}, {Poir{\`e}}, {Popa}, {Pradier}, {Probst},
  {Pulvirenti}, {Qu{\'e}m{\'e}ner}, {Randazzo}, {Razzaque}, {Real}, {Reck},
  {Riccobene}, {Romanov}, \& {Rovelli}}]{2022EPJC...82..317A}
{Aiello}, S., {Albert}, A., {Alshamsi}, M., {et~al.} 2022, European Physical
  Journal C, 82, 317, \dodoi{10.1140/epjc/s10052-022-10137-y}

\bibitem[{{Aiello} {et~al.}(2024){Aiello}, {Albert}, {Alshamsi}, {Alves Garre},
  {Aly}, {Ambrosone}, {Ameli}, {Andre}, {Androutsou}, {Anguita}, {Aphecetche},
  {Ardid}, {Ardid}, {Atmani}, {Aublin}, {Badaracco}, {Bailly-Salins},
  {Barda{\v{c}}ov{\'a}}, {Baret}, {Bariego-Quintana}, {Basegmez du Pree},
  {Becherini}, {Bendahman}, {Benfenati}, {Benhassi}, {Benoit}, {Berbee},
  {Bertin}, {Biagi}, {Boettcher}, {Bonanno}, {Boumaaza}, {Bouta}, {Bouwhuis},
  {Bozza}, {Bozza}, {Br{\^a}nza{\c{s}}}, {Bretaudeau}, {Breuhaus}, {Bruijn},
  {Brunner}, {Bruno}, {Buis}, {Buompane}, {Busto}, {Caiffi}, {Calvo},
  {Campion}, {Capone}, {Carenini}, {Carretero}, {Cartraud}, {Castaldi},
  {Cecchini}, {Celli}, {Cerisy}, {Chabab}, {Chadolias}, {Chen}, {Cherubini},
  {Chiarusi}, {Circella}, {Cocimano}, {Coelho}, {Coleiro}, {Coniglione},
  {Coyle}, {Creusot}, {Cuttone}, {Dallier}, {Darras}, {De Benedittis}, {De
  Martino}, {Decoene}, {Del Burgo}, {Del Rosso}, {Di Mauro}, {Di Palma},
  {D{\'\i}az}, {Diaz}, {Diego-Tortosa}, {Distefano}, {Domi}, {Donzaud},
  {Dornic}, {D{\"o}rr}, {Drakopoulou}, {Drouhin}, {Dvornick{\'y}}, {Eberl},
  {Eckerov{\'a}}, {Eddymaoui}, {van Eeden}, {Eff}, {van Eijk}, {El Bojaddaini},
  {El Hedri}, {Enzenh{\"o}fer}, {Ferrara}, {Filipovi{\'c}}, {Filippini},
  {Franciotti}, {Fusco}, {Gabriel}, {Gagliardini}, {Gal}, {Garc{\'\i}a
  M{\'e}ndez}, {Garcia Soto}, {Gatius Oliver}, {Gei{\ss}elbrecht}, {Ghaddari},
  {Gialanella}, {Gibson}, {Giorgio}, {Goos}, {Goswami}, {Goupilliere},
  {Gozzini}, {Gracia}, {Graf}, {Guidi}, {Guillon}, {Guti{\'e}rrez}, {van
  Haren}, {Heijboer}, {Hekalo}, {Hennig}, {Hern{\'a}ndez-Rey}, {Ibnsalih},
  {Illuminati}, {de Jong}, {de Jong}, {Jung}, {Kalaczy{\'n}ski}, {Kalekin},
  {Katz}, {Kistauri}, {Kopper}, {Kouchner}, {Kueviakoe}, {Kulikovskiy},
  {Kvatadze}, {Labalme}, {Lahmann}, {Larosa}, {Lastoria}, {Lazo}, {Le Stum},
  {Lehaut}, {Leonora}, {Lessing}, {Levi}, {Clark}, {Longhitano}, {Magnani},
  {Majumdar}, {Malerba}, {Mamedov}, {Ma{\'n}czak}, {Manfreda}, {Manzaneda},
  {Marconi}, {Margiotta}, {Marinelli}, {Markou}, {Martin},
  {Mart{\'\i}nez-Mora}, {Marzaioli}, {Mastrodicasa}, {Mastroianni},
  {Miccich{\`e}}, {Miele}, {Migliozzi}, {Migneco}, {Mitsou}, {Mollo},
  {Morales-Gallegos}, {Morga}, {Moussa}, {Mateo}, {Muller}, {Musone},
  {Musumeci}, {Navas}, {Nayerhoda}, {Nicolau}, {Nkosi}, {Fearraigh},
  {Oliviero}, {Orlando}, {Oukacha}, {Paesani}, {Palacios Gonz{\'a}lez},
  {Papalashvili}, {Parisi}, {Gomez}, {P{\u{a}}un}, {P{\u{a}}v{\u{a}}la{\c{s}}},
  {Pe{\~n}a Mart{\'\i}nez}, \& {Perrin-Terrin}}]{2024APh...16202990A}
---. 2024, Astroparticle Physics, 162, 102990,
  \dodoi{10.1016/j.astropartphys.2024.102990}

\bibitem[{{Albert} {et~al.}(2024){Albert}, {Alves}, {Andr{\'e}}, {Ardid},
  {Ardid}, {Aubert}, {Aublin}, {Baret}, {Basa}, {Becherini}, {Belhorma},
  {Bendahman}, {Benfenati}, {Bertin}, {Biagi}, {Bissinger}, {Boumaaza},
  {Bouta}, {Bouwhuis}, {Br{\^a}nza{\c{s}}}, {Bruijn}, {Brunner}, {Busto},
  {Caiffi}, {Calvo}, {Campion}, {Capone}, {Caramete}, {Carenini}, {Carr},
  {Carretero}, {Celli}, {Cerisy}, {Chabab}, {Cherkaoui El Moursli}, {Chiarusi},
  {Circella}, {Coelho}, {Coleiro}, {Coniglione}, {Coyle}, {Creusot}, {Cruz},
  {D{\'\i}az}, {de Martino}, {Distefano}, {di Palma}, {Domi}, {Donzaud},
  {Dornic}, {Drouhin}, {Eberl}, {van Eeden}, {van Eijk}, {El Hedri}, {El
  Khayati}, {Enzenh{\"o}fer}, {Fermani}, {Ferrara}, {Filippini}, {Fusco},
  {Gagliardini}, {Garc{\'\i}a}, {Gatius Oliver}, {Gay}, {Gei{\ss}elbrecht},
  {Glotin}, {Gozzini}, {Gracia Ruiz}, {Graf}, {Guidi}, {Haegel}, {Hallmann},
  {van Haren}, {Heijboer}, {Hello}, {Hern{\'a}ndez-Rey}, {H{\"o}{\ss}l},
  {Hofest{\"a}dt}, {Huang}, {Illuminati}, {James}, {Jisse-Jung}, {de Jong}, {de
  Jong}, {Kadler}, {Kalekin}, {Katz}, {Kouchner}, {Kovalev}, {Kovalev},
  {Kreykenbohm}, {Kulikovskiy}, {Lahmann}, {Lamoureux}, {Lazo}, {Lef{\`e}vre},
  {Leonora}, {Levi}, {Le Stum}, {Lopez-Coto}, {Loucatos}, {Maderer}, {Manczak},
  {Marcelin}, {Margiotta}, {Marinelli}, {Mart{\'\i}nez-Mora}, {Migliozzi},
  {Moussa}, {Muller}, {Navas}, {Nezri}, {Fearraigh}, {Oukacha}, {P{\u{a}}un},
  {P{\u{a}}v{\u{a}}la{\c{s}}}, {Pe{\~n}a-Mart{\'\i}nez}, {Perrin-Terrin},
  {Pestel}, {Piattelli}, {Plavin}, {Poir{\`e}}, {Popa}, {Pradier}, {Pushkarev},
  {Randazzo}, {Real}, {Reck}, {Riccobene}, {Romanov}, {S{\'a}nchez-Losa},
  {Saina}, {Salesa Greus}, {Samtleben}, {Sanguineti}, {Sapienza}, {Schnabel},
  {Schumann}, {Sch{\"u}ssler}, {Seneca}, {Spurio}, {Stolarczyk}, {Taiuti},
  {Tayalati}, {Tingay}, {Troitsky}, {Vallage}, {Vannoye}, {van Elewyck},
  {Viola}, {Vivolo}, {Wilms}, {Zavatarelli}, {Zegarelli}, {Zornoza},
  {Z{\'u}{\~n}iga}, {(ANTARES Collaboration)}, {Hovatta}, {Kiehlmann},
  {Liodakis}, {Pavlidou}, {Readhead}, \& {(Ovro
  Collaboration)}}]{2024ApJ...964....3A}
{Albert}, A., {Alves}, S., {Andr{\'e}}, M., {et~al.} 2024, \apj, 964, 3,
  \dodoi{10.3847/1538-4357/ad1f5b}

\bibitem[{Amelino-Camelia {et~al.}(2025)Amelino-Camelia, D'Amico, Fabiano,
  Frattulillo, Gubitosi, Moia, \& Rosati}]{grb}
Amelino-Camelia, G., D'Amico, G., Fabiano, G., {et~al.} 2025, On testing
  in-vacuo dispersion with the most energetic neutrinos: KM3-230213A case
  study.
\newblock \doarXiv{2502.13093}

\bibitem[{{Ando} {et~al.}(2023){Ando}, {Ekanger}, {Horiuchi}, \&
  {Koshio}}]{2023PJAB...99..460A}
{Ando}, S., {Ekanger}, N., {Horiuchi}, S., \& {Koshio}, Y. 2023, Proceedings of
  the Japan Academy, Series B, 99, 460, \dodoi{10.2183/pjab.99.026}

\bibitem[{{Ball} {et~al.}(2023){Ball}, {Kothes}, {Rosolowsky}, {West},
  {Becker}, {Filipovi{\'c}}, {Gaensler}, {Hopkins}, {Koribalski}, {Landecker},
  {Leahy}, {Marvil}, {Sun}, {Bufano}, {Carretti}, {Ingallinera}, {Van Eck}, \&
  {Willis}}]{Ball2023}
{Ball}, B.~D., {Kothes}, R., {Rosolowsky}, E., {et~al.} 2023, \mnras, 524,
  1396, \dodoi{10.1093/mnras/stad1953}

\bibitem[{{Bechtol} {et~al.}(2017){Bechtol}, {Ahlers}, {Di Mauro}, {Ajello}, \&
  {Vandenbroucke}}]{2017ApJ...836...47B}
{Bechtol}, K., {Ahlers}, M., {Di Mauro}, M., {Ajello}, M., \& {Vandenbroucke},
  J. 2017, \apj, 836, 47, \dodoi{10.3847/1538-4357/836/1/47}

\bibitem[{{Bilicki} {et~al.}(2014){Bilicki}, {Jarrett}, {Peacock}, {Cluver}, \&
  {Steward}}]{2014ApJS..210....9B}
{Bilicki}, M., {Jarrett}, T.~H., {Peacock}, J.~A., {Cluver}, M.~E., \&
  {Steward}, L. 2014, \apjs, 210, 9, \dodoi{10.1088/0067-0049/210/1/9}

\bibitem[{{Bradley} {et~al.}(2025){Bradley}, {Smeaton}, {Tothill},
  {Filipovi{\'c}}, {Becker}, {Hopkins}, {Koribalski}, {Lazarevi{\'c}}, {Leahy},
  {Rowell}, {Velovi{\'c}}, \& {Uro{\v{s}}evi{\'c}}}]{2025PASA...42...32B}
{Bradley}, A.~C., {Smeaton}, Z.~J., {Tothill}, N.~F.~H., {et~al.} 2025, \pasa,
  e032, \dodoi{10.1017/pasa.2025.15}

\bibitem[{{Bykov} {et~al.}(2015){Bykov}, {Ellison}, {Gladilin}, \&
  {Osipov}}]{2015MNRAS.453..113B}
{Bykov}, A.~M., {Ellison}, D.~C., {Gladilin}, P.~E., \& {Osipov}, S.~M. 2015,
  \mnras, 453, 113, \dodoi{10.1093/mnras/stv1606}

\bibitem[{{Caprioli}(2015)}]{2015ApJ...811L..38C}
{Caprioli}, D. 2015, \apjl, 811, L38, \dodoi{10.1088/2041-8205/811/2/L38}

\bibitem[{{Collier} {et~al.}(2018){Collier}, {Tingay}, {Callingham}, {Norris},
  {Filipovi{\'c}}, {Galvin}, {Huynh}, {Intema}, {Marvil}, {O'Brien}, {Roper},
  {Sirothia}, {Tothill}, {Bell}, {For}, {Gaensler}, {Hancock}, {Hindson},
  {Hurley-Walker}, {Johnston-Hollitt}, {Kapi{\'n}ska}, {Lenc}, {Morgan},
  {Procopio}, {Staveley-Smith}, {Wayth}, {Wu}, {Zheng}, {Heywood}, \&
  {Popping}}]{2018MNRAS.477..578C}
{Collier}, J.~D., {Tingay}, S.~J., {Callingham}, J.~R., {et~al.} 2018, \mnras,
  477, 578, \dodoi{10.1093/mnras/sty564}

\bibitem[{{Condon} {et~al.}(1998){Condon}, {Cotton}, {Greisen}, {Yin},
  {Perley}, {Taylor}, \& {Broderick}}]{1998AJ....115.1693C}
{Condon}, J.~J., {Cotton}, W.~D., {Greisen}, E.~W., {et~al.} 1998, \aj, 115,
  1693, \dodoi{10.1086/300337}

\bibitem[{{Condon} {et~al.}(2021){Condon}, {Cotton}, {Jarrett}, {Marchetti},
  {Matthews}, {Mauch}, \& {Moloko}}]{Condon2021}
{Condon}, J.~J., {Cotton}, W.~D., {Jarrett}, T., {et~al.} 2021, \apjs, 257, 35,
  \dodoi{10.3847/1538-4365/ac1f17}

\bibitem[{{Cotton} {et~al.}(2024){Cotton}, {Filipovi{\'c}}, {Camilo},
  {Indebetouw}, {Alsaberi}, {Anih}, {Baker}, {Bastian}, {Boji{\v{c}}i{\'c}},
  {Carli}, {Cavallaro}, {Crawford}, {Dai}, {Haberl}, {Levin}, {Luken},
  {Pennock}, {Rajabpour}, {Stappers}, {van Loon}, {Zijlstra}, {Buchner},
  {Geyer}, {Goedhart}, \& {Serylak}}]{2024MNRAS.529.2443C}
{Cotton}, W.~D., {Filipovi{\'c}}, M.~D., {Camilo}, F., {et~al.} 2024, \mnras,
  529, 2443, \dodoi{10.1093/mnras/stae277}

\bibitem[{{Cutri} {et~al.}(2013){Cutri}, {Wright}, {Conrow}, {Fowler},
  {Eisenhardt}, {Grillmair}, {Kirkpatrick}, {Masci}, {McCallon}, {Wheelock},
  {Fajardo-Acosta}, {Yan}, {Benford}, {Harbut}, {Jarrett}, {Lake}, {Leisawitz},
  {Ressler}, {Stanford}, {Tsai}, {Liu}, {Helou}, {Mainzer}, {Gettings},
  {Gonzalez}, {Hoffman}, {Marsh}, {Padgett}, {Skrutskie}, {Beck}, {Papin}, \&
  {Wittman}}]{2013wise.rept....1C}
{Cutri}, R.~M., {Wright}, E.~L., {Conrow}, T., {et~al.} 2013, {Explanatory
  Supplement to the AllWISE Data Release Products}, Explanatory Supplement to
  the AllWISE Data Release Products, by R. M. Cutri et al.

\bibitem[{{Driessen} {et~al.}(2024){Driessen}, {Pritchard}, {Murphy}, {Heald},
  {Robrade}, {Das}, {Duchesne}, {Kaplan}, {Lenc}, {Lynch}, {Mitchell-Bolton},
  {Pope}, {Rose}, {Stelzer}, {Wang}, \& {Zic}}]{2024PASA...41...84D}
{Driessen}, L.~N., {Pritchard}, J., {Murphy}, T., {et~al.} 2024, \pasa, 41,
  e084, \dodoi{10.1017/pasa.2024.72}

\bibitem[{{Duric}(2000)}]{2000immm.proc..179D}
{Duric}, N. 2000, in Proceedings 232. WE-Heraeus Seminar, ed. E.~M.
  {Berkhuijsen}, R.~{Beck}, \& R.~A.~M. {Walterbos}, 179--186

\bibitem[{{Duric} {et~al.}(1995){Duric}, {Gordon}, {Goss}, {Viallefond}, \&
  {Lacey}}]{1995ApJ...445..173D}
{Duric}, N., {Gordon}, S.~M., {Goss}, W.~M., {Viallefond}, F., \& {Lacey}, C.
  1995, \apj, 445, 173, \dodoi{10.1086/175683}

\bibitem[{{Fang} {et~al.}(2020){Fang}, {Banerjee}, {Charles}, \&
  {Omori}}]{2020ApJ...894..112F}
{Fang}, K., {Banerjee}, A., {Charles}, E., \& {Omori}, Y. 2020, \apj, 894, 112,
  \dodoi{10.3847/1538-4357/ab8561}

\bibitem[{Filipovi{\'c} \& Tothill(2021)}]{book1}
Filipovi{\'c}, M.~D., \& Tothill, N. F.~H. 2021, Principles of Multimessenger
  Astronomy, 2514-3433 (IOP Publishing), \dodoi{10.1088/2514-3433/ac087e}

\bibitem[{{Filipovi{\'c}} {et~al.}(2021){Filipovi{\'c}}, {Boji{\v{c}}i{\'c}},
  {Grieve}, {Norris}, {Tothill}, {Shobhana}, {Rudnick}, {Prandoni},
  {Andernach}, {Hurley-Walker}, {Alsaberi}, {Anderson}, {Collier}, {Crawford},
  {For}, {Galvin}, {Haberl}, {Hopkins}, {Ingallinera}, {Kavanagh},
  {Koribalski}, {Kothes}, {Leahy}, {Leverenz}, {Maggi}, {Maitra}, {Marvil},
  {Pannuti}, {Park}, {Payne}, {Pennock}, {Riggi}, {Rowell}, {Sano}, {Sasaki},
  {Staveley-Smith}, {Trigilio}, {Umana}, {Uro{\v{s}}evi{\'c}}, {van Loon}, \&
  {Vardoulaki}}]{2021MNRAS.507.2885F}
{Filipovi{\'c}}, M.~D., {Boji{\v{c}}i{\'c}}, I.~S., {Grieve}, K.~R., {et~al.}
  2021, \mnras, 507, 2885, \dodoi{10.1093/mnras/stab2249}

\bibitem[{{Filipovi{\'c}} {et~al.}(2022){Filipovi{\'c}}, {Payne}, {Alsaberi},
  {Norris}, {Macgregor}, {Rudnick}, {Koribalski}, {Leahy}, {Ducci}, {Kothes},
  {Andernach}, {Barnes}, {Boji{\v{c}}i{\'c}}, {Bozzetto}, {Brose}, {Collier},
  {Crawford}, {Crocker}, {Dai}, {Galvin}, {Haberl}, {Heber}, {Hill}, {Hopkins},
  {Hurley-Walker}, {Ingallinera}, {Jarrett}, {Kavanagh}, {Lenc}, {Luken},
  {Mackey}, {Manojlovi{\'c}}, {Maggi}, {Maitra}, {Pennock}, {Points}, {Riggi},
  {Rowell}, {Safi-Harb}, {Sano}, {Sasaki}, {Shabala}, {Stevens}, {van Loon},
  {Tothill}, {Umana}, {Uro{\v{s}}evi{\'c}}, {Velovi{\'c}}, {Vernstrom}, {West},
  \& {Wan}}]{2022MNRAS.512..265F}
{Filipovi{\'c}}, M.~D., {Payne}, J.~L., {Alsaberi}, R.~Z.~E., {et~al.} 2022,
  \mnras, 512, 265, \dodoi{10.1093/mnras/stac210}

\bibitem[{{Filipovi{\'c}} {et~al.}(2024){Filipovi{\'c}}, {Lazarevi{\'c}},
  {Araya}, {Hurley-Walker}, {Kothes}, {Sano}, {Rowell}, {Martin}, {Fukui},
  {Alsaberi}, {Arbutina}, {Ball}, {Bordiu}, {Brose}, {Bufano},
  {Burger-Scheidlin}, {Anne Collins}, {Crawford}, {Dai}, {William Duchesne},
  {Fuller}, {Hopkins}, {Ingallinera}, {Inoue}, {Jarrett}, {Silvia Koribalski},
  {Leahy}, {Luken}, {Mackey}, {Macgregor}, {Norris}, {Payne}, {Riggi},
  {Riseley}, {Sasaki}, {Smeaton}, {Sushch}, {Stupar}, {Umana},
  {Uro{\v{s}}evi{\'c}}, {Velovi{\'c}}, {Vernstrom}, {Vukoti{\'c}}, \&
  {West}}]{2024PASA...41..112F}
{Filipovi{\'c}}, M.~D., {Lazarevi{\'c}}, S., {Araya}, M., {et~al.} 2024, \pasa,
  41, e112, \dodoi{10.1017/pasa.2024.93}

\bibitem[{Filipović \& Tothill(2021)}]{book2}
Filipović, M.~D., \& Tothill, N. F.~H., eds. 2021, Multimessenger Astronomy in
  Practice (IOP Publishing)

\bibitem[{{Gaensler} {et~al.}(2010){Gaensler}, {Landecker}, {Taylor}, \&
  {POSSUM Collaboration}}]{Gaensler2010}
{Gaensler}, B.~M., {Landecker}, T.~L., {Taylor}, A.~R., \& {POSSUM
  Collaboration}. 2010, in American Astronomical Society Meeting Abstracts,
  Vol. 215, American Astronomical Society Meeting Abstracts \#215, 470.13

\bibitem[{{Gordon} {et~al.}(2021){Gordon}, {Boyce}, {O'Dea}, {Rudnick},
  {Andernach}, {Vantyghem}, {Baum}, {Bui}, {Dionyssiou}, {Safi-Harb}, \&
  {Sander}}]{2021ApJS..255...30G}
{Gordon}, Y.~A., {Boyce}, M.~M., {O'Dea}, C.~P., {et~al.} 2021, \apjs, 255, 30,
  \dodoi{10.3847/1538-4365/ac05c0}

\bibitem[{{Griffith} {et~al.}(1995){Griffith}, {Wright}, {Burke}, \&
  {Ekers}}]{1995ApJS...97..347G}
{Griffith}, M.~R., {Wright}, A.~E., {Burke}, B.~F., \& {Ekers}, R.~D. 1995,
  \apjs, 97, 347, \dodoi{10.1086/192146}

\bibitem[{{Guzman} {et~al.}(2019){Guzman}, {Whiting}, {Voronkov}, {Mitchell},
  {Ord}, {Collins}, {Marquarding}, {Lahur}, {Maher}, {Van Diepen}, {Bannister},
  {Wu}, {Lenc}, {Khoo}, \& {Bastholm}}]{2019ascl.soft12003G}
{Guzman}, J., {Whiting}, M., {Voronkov}, M., {et~al.} 2019, {ASKAPsoft: ASKAP
  science data processor software}, Astrophysics Source Code Library, record
  ascl:1912.003

\bibitem[{{H.~E.~S.~S. Collaboration} {et~al.}(2024){H.~E.~S.~S.
  Collaboration}, {Aharonian}, {Ait Benkhali}, {Aschersleben}, {Ashkar},
  {Backes}, {Barbosa Martins}, {Batzofin}, {Becherini}, {Berge},
  {Bernl{\"o}hr}, {Bi}, {B{\"o}ttcher}, {Boisson}, {Bolmont}, {de Lavergne},
  {Borowska}, {Bouyahiaoui}, {Breuhaus}, {Brose}, {Brown}, {Brun}, {Bruno},
  {Bulik}, {Burger-Scheidlin}, {Caroff}, {Casanova}, {Cecil}, {Celic},
  {Cerruti}, {Chand}, {Chandra}, {Chen}, {Chibueze}, {Chibueze}, {Cotter},
  {Dai}, {Mbarubucyeye}, {Djannati-Ata{\"\i}}, {Dmytriiev}, {Doroshenko},
  {Egberts}, {Einecke}, {Ernenwein}, {Filipovic}, {Fontaine},
  {F{\"u}{\ss}ling}, {Funk}, {Gabici}, {Ghafourizadeh}, {Giavitto}, {Glawion},
  {Glicenstein}, {Grolleron}, {Haerer}, {Hinton}, {Hofmann}, {Holch}, {Holler},
  {Horns}, {Jamrozy}, {Jankowsky}, {Jardin-Blicq}, {Joshi}, {Jung-Richardt},
  {Kasai}, {Katarzy{\'n}ski}, {Khatoon}, {Kh{\'e}lifi}, {Klepser},
  {Klu{\'z}niak}, {Komin}, {Kosack}, {Kostunin}, {Kundu}, {Lang}, {Le Stum},
  {Leitl}, {Lemi{\`e}re}, {Lenain}, {Leuschner}, {Lohse}, {Luashvili},
  {Lypova}, {Mackey}, {Malyshev}, {Malyshev}, {Marandon}, {Marchegiani},
  {Marcowith}, {Mart{\'\i}-Devesa}, {Marx}, {Mehta}, {Mitchell}, {Moderski},
  {Mohrmann}, {Montanari}, {Moulin}, {Murach}, {Nakashima}, {de Naurois},
  {Niemiec}, {Noel}, {Ohm}, {Olivera-Nieto}, {de Ona Wilhelmi}, {Ostrowski},
  {Panny}, {Panter}, {Parsons}, {Peron}, {Prokhorov}, {P{\"u}hlhofer}, {Punch},
  {Quirrenbach}, {Reichherzer}, {Reimer}, {Reimer}, {Ren}, {Renaud}, {Reville},
  {Rieger}, {Rowell}, {Rudak}, {Ricarte}, {Ruiz-Velasco}, {Sahakian},
  {Salzmann}, {Santangelo}, {Sasaki}, {Sch{\"a}fer}, {Sch{\"u}ssler},
  {Schwanke}, {Shapopi}, {Sol}, {Specovius}, {Spencer}, {Stawarz}, {Steenkamp},
  {Steinmassl}, {Steppa}, {Streil}, {Sushch}, {Suzuki}, {Takahashi}, {Tanaka},
  {Taylor}, {Terrier}, {Tsirou}, {Tsuji}, {Unbehaun}, {van Eldik}, {Vecchi},
  {Veh}, {Venter}, {Vink}, {Wach}, {Wagner}, {Werner}, {White}, {Wierzcholska},
  {Wong}, {Zacharias}, {Zargaryan}, {Zdziarski}, {Zech}, {Zouari}, \&
  {{\.Z}ywucka}}]{2024Sci...383..402H}
{H.~E.~S.~S. Collaboration}, {Aharonian}, F., {Ait Benkhali}, F., {et~al.}
  2024, Science, 383, 402, \dodoi{10.1126/science.adi2048}

\bibitem[{{Hajela} {et~al.}(2019){Hajela}, {Mooley}, {Intema}, \&
  {Frail}}]{2019MNRAS.490.4898H}
{Hajela}, A., {Mooley}, K.~P., {Intema}, H.~T., \& {Frail}, D.~A. 2019, \mnras,
  490, 4898, \dodoi{10.1093/mnras/stz2918}

\bibitem[{{Hale} {et~al.}(2021){Hale}, {McConnell}, {Thomson}, {Lenc}, {Heald},
  {Hotan}, {Leung}, {Moss}, {Murphy}, {Pritchard}, {Sadler}, {Stewart}, \&
  {Whiting}}]{2021PASA...38...58H}
{Hale}, C.~L., {McConnell}, D., {Thomson}, A.~J.~M., {et~al.} 2021, \pasa, 38,
  e058, \dodoi{10.1017/pasa.2021.47}

\bibitem[{{Hancock} {et~al.}(2018){Hancock}, {Trott}, \&
  {Hurley-Walker}}]{hancock2018}
{Hancock}, P.~J., {Trott}, C.~M., \& {Hurley-Walker}, N. 2018, PASA, 35, e011,
  \dodoi{10.1017/pasa.2018.3}

\bibitem[{{Harvey-Smith} {et~al.}(2010){Harvey-Smith}, {Gaensler}, {Kothes},
  {Townsend}, {Heald}, {Ng}, \& {Green}}]{2010ApJ...712.1157H}
{Harvey-Smith}, L., {Gaensler}, B.~M., {Kothes}, R., {et~al.} 2010, \apj, 712,
  1157, \dodoi{10.1088/0004-637X/712/2/1157}

\bibitem[{{Haubner} {et~al.}(2025){Haubner}, {Sasaki}, {Mitchell}, {Ponti},
  {Rowell}, {Einecke}, {Filipovi{\'c}}, {Lazarevi{\'c}}, {P{\"u}hlhofer}, \&
  {Strong}}]{2025arXiv250112990H}
{Haubner}, K., {Sasaki}, M., {Mitchell}, A., {et~al.} 2025, arXiv e-prints,
  arXiv:2501.12990, \dodoi{10.48550/arXiv.2501.12990}

\bibitem[{{Hirata} {et~al.}(1987){Hirata}, {Kajita}, {Koshiba}, {Nakahata},
  {Oyama}, {Sato}, {Suzuki}, {Takita}, {Totsuka}, {Kifune}, {Suda},
  {Takahashi}, {Tanimori}, {Miyano}, {Yamada}, {Beier}, {Feldscher}, {Kim},
  {Mann}, {Newcomer}, {van}, {Zhang}, \& {Cortez}}]{1987PhRvL..58.1490H}
{Hirata}, K., {Kajita}, T., {Koshiba}, M., {et~al.} 1987, \prl, 58, 1490,
  \dodoi{10.1103/PhysRevLett.58.1490}

\bibitem[{{Hotan} {et~al.}(2021){Hotan}, {Bunton}, {Chippendale}, {Whiting},
  {Tuthill}, {Moss}, {McConnell}, {Amy}, {Huynh}, {Allison}, {Anderson},
  {Bannister}, {Bastholm}, {Beresford}, {Bock}, {Bolton}, {Chapman}, {Chow},
  {Collier}, {Cooray}, {Cornwell}, {Diamond}, {Edwards}, {Feain}, {Franzen},
  {George}, {Gupta}, {Hampson}, {Harvey-Smith}, {Hayman}, {Heywood}, {Jacka},
  {Jackson}, {Jackson}, {Jeganathan}, {Johnston}, {Kesteven}, {Kleiner},
  {Koribalski}, {Lee-Waddell}, {Lenc}, {Lensson}, {Mackay}, {Mahony},
  {McClure-Griffiths}, {McConigley}, {Mirtschin}, {Ng}, {Norris}, {Pearce},
  {Phillips}, {Pilawa}, {Raja}, {Reynolds}, {Roberts}, {Roxby}, {Sadler},
  {Shields}, {Schinckel}, {Serra}, {Shaw}, {Sweetnam}, {Troup}, {Tzioumis},
  {Voronkov}, \& {Westmeier}}]{2021PASA...38....9H}
{Hotan}, A.~W., {Bunton}, J.~D., {Chippendale}, A.~P., {et~al.} 2021, \pasa,
  38, e009

\bibitem[{{Hutschenreuter} {et~al.}(2022){Hutschenreuter}, {Anderson}, {Betti},
  {Bower}, {Brown}, {Br{\"u}ggen}, {Carretti}, {Clarke}, {Clegg}, {Costa},
  {Croft}, {Van Eck}, {Gaensler}, {de Gasperin}, {Haverkorn}, {Heald}, {Hull},
  {Inoue}, {Johnston-Hollitt}, {Kaczmarek}, {Law}, {Ma}, {MacMahon}, {Mao},
  {Riseley}, {Roy}, {Shanahan}, {Shimwell}, {Stil}, {Sobey}, {O'Sullivan},
  {Tasse}, {Vacca}, {Vernstrom}, {Williams}, {Wright}, \&
  {En{\ss}lin}}]{2022A&A...657A..43H}
{Hutschenreuter}, S., {Anderson}, C.~S., {Betti}, S., {et~al.} 2022, \aap, 657,
  A43, \dodoi{10.1051/0004-6361/202140486}

\bibitem[{{IceCube Collaboration} {et~al.}(2018){IceCube Collaboration},
  {Aartsen}, {Ackermann}, {Adams}, {Aguilar}, {Ahlers}, {Ahrens}, {Al Samarai},
  {Altmann}, {Andeen}, {Anderson}, {Ansseau}, {Anton}, {Arg{\"u}elles},
  {Auffenberg}, {Axani}, {Bagherpour}, {Bai}, {Barron}, {Barwick}, {Baum},
  {Bay}, {Beatty}, {Becker Tjus}, {Becker}, {BenZvi}, {Berley}, {Bernardini},
  {Besson}, {Binder}, {Bindig}, {Blaufuss}, {Blot}, {Bohm}, {B{\"o}rner},
  {Bos}, {B{\"o}ser}, {Botner}, {Bourbeau}, {Bourbeau}, {Bradascio}, {Braun},
  {Brenzke}, {Bretz}, {Bron}, {Brostean-Kaiser}, {Burgman}, {Busse}, {Carver},
  {Cheung}, {Chirkin}, {Christov}, {Clark}, {Classen}, {Coenders}, {Collin},
  {Conrad}, {Coppin}, {Correa}, {Cowen}, {Cross}, {Dave}, {Day}, {de
  Andr{\'e}}, {De Clercq}, {DeLaunay}, {Dembinski}, {De Ridder}, {Desiati}, {de
  Vries}, {de Wasseige}, {de With}, {DeYoung}, {D{\'\i}az-V{\'e}lez}, {di
  Lorenzo}, {Dujmovic}, {Dumm}, {Dunkman}, {Dvorak}, {Eberhardt}, {Ehrhardt},
  {Eichmann}, {Eller}, {Evenson}, {Fahey}, {Fazely}, {Felde}, {Filimonov},
  {Finley}, {Flis}, {Franckowiak}, {Friedman}, {Fritz}, {Gaisser}, {Gallagher},
  {Gerhardt}, {Ghorbani}, {Glauch}, {Gl{\"u}senkamp}, {Goldschmidt},
  {Gonzalez}, {Grant}, {Griffith}, {Haack}, {Hallgren}, {Halzen}, {Hanson},
  {Hebecker}, {Heereman}, {Helbing}, {Hellauer}, {Hickford}, {Hignight},
  {Hill}, {Hoffman}, {Hoffmann}, {Hoinka}, {Hokanson-Fasig}, {Hoshina},
  {Huang}, {Huber}, {Hultqvist}, {H{\"u}nnefeld}, {Hussain}, {In}, {Iovine},
  {Ishihara}, {Jacobi}, {Japaridze}, {Jeong}, {Jero}, {Jones}, {Kalaczynski},
  {Kang}, {Kappes}, {Kappesser}, {Karg}, {Karle}, {Katz}, {Kauer}, {Keivani},
  {Kelley}, {Kheirandish}, {Kim}, {Kim}, {Kintscher}, {Kiryluk}, {Kittler},
  {Klein}, {Koirala}, {Kolanoski}, {K{\"o}pke}, {Kopper}, {Kopper},
  {Koschinsky}, {Koskinen}, {Kowalski}, {Krings}, {Kroll}, {Kr{\"u}ckl},
  {Kunwar}, {Kurahashi}, {Kuwabara}, {Kyriacou}, {Labare}, {Lanfranchi},
  {Larson}, {Lauber}, {Leonard}, {Lesiak-Bzdak}, {Leuermann}, {Liu}, {Lozano
  Mariscal}, {Lu}, {L{\"u}nemann}, {Luszczak}, {Madsen}, {Maggi}, {Mahn},
  {Mancina}, {Maruyama}, {Mase}, {Maunu}, {Meagher}, {Medici}, {Meier},
  {Menne}, {Merino}, {Meures}, {Miarecki}, {Micallef}, {Moment{\'e}},
  {Montaruli}, {Moore}, {Morse}, {Moulai}, {Nahnhauer}, {Nakarmi}, {Naumann},
  \& {Neer}}]{2018Sci...361.1378I}
{IceCube Collaboration}, {Aartsen}, M.~G., {Ackermann}, M., {et~al.} 2018,
  Science, 361, eaat1378, \dodoi{10.1126/science.aat1378}

\bibitem[{{IceCube Collaboration} {et~al.}(2022){IceCube Collaboration},
  {Abbasi}, {Ackermann}, {Adams}, {Aguilar}, {Ahlers}, {Ahrens}, {Alameddine},
  {Alispach}, {Alves}, {Amin}, {Andeen}, {Anderson}, {Anton}, {Arg{\"u}elles},
  {Ashida}, {Axani}, {Bai}, {Balagopal}, {Barbano}, {Barwick}, {Bastian},
  {Basu}, {Baur}, {Bay}, {Beatty}, {Becker}, {Becker Tjus}, {Bellenghi},
  {Benzvi}, {Berley}, {Bernardini}, {Besson}, {Binder}, {Bindig}, {Blaufuss},
  {Blot}, {Boddenberg}, {Bontempo}, {Borowka}, {B{\"o}ser}, {Botner},
  {B{\"o}ttcher}, {Bourbeau}, {Bradascio}, {Braun}, {Brinson}, {Bron},
  {Brostean-Kaiser}, {Browne}, {Burgman}, {Burley}, {Busse}, {Campana},
  {Carnie-Bronca}, {Chen}, {Chen}, {Chirkin}, {Choi}, {Clark}, {Clark},
  {Classen}, {Coleman}, {Collin}, {Conrad}, {Coppin}, {Correa}, {Cowen},
  {Cross}, {Dappen}, {Dave}, {de Clercq}, {Delaunay}, {Delgado L{\'o}pez},
  {Dembinski}, {Deoskar}, {Desai}, {Desiati}, {de Vries}, {de Wasseige}, {de
  With}, {Deyoung}, {Diaz}, {D{\'\i}az-V{\'e}lez}, {Dittmer}, {Dujmovic},
  {Dunkman}, {Duvernois}, {Dvorak}, {Ehrhardt}, {Eller}, {Engel}, {Erpenbeck},
  {Evans}, {Evenson}, {Fan}, {Fazely}, {Fedynitch}, {Feigl}, {Fiedlschuster},
  {Fienberg}, {Filimonov}, {Finley}, {Fischer}, {Fox}, {Franckowiak},
  {Friedman}, {Fritz}, {F{\"u}rst}, {Gaisser}, {Gallagher}, {Ganster},
  {Garcia}, {Garrappa}, {Gerhardt}, {Ghadimi}, {Glaser}, {Glauch},
  {Gl{\"u}senkamp}, {Goldschmidt}, {Gonzalez}, {Goswami}, {Grant},
  {Gr{\'e}goire}, {Griswold}, {G{\"u}nther}, {Gutjahr}, {Haack}, {Hallgren},
  {Halliday}, {Halve}, {Halzen}, {Hanson}, {Hardin}, {Harnisch}, {Haungs},
  {Hebecker}, {Helbing}, {Henningsen}, {Hettinger}, {Hickford}, {Hignight},
  {Hill}, {Hill}, {Hoffman}, {Hoffmann}, {Hokanson-Fasig}, {Hoshina}, {Huang},
  {Huber}, {Huber}, {Hultqvist}, {H{\"u}nnefeld}, {Hussain}, {Hymon}, {in},
  {Iovine}, {Ishihara}, {Jansson}, {Japaridze}, {Jeong}, {Jin}, {Jones},
  {Kang}, {Kang}, {Kang}, {Kappes}, {Kappesser}, {Kardum}, {Karg}, {Karl},
  {Karle}, {Katz}, {Kauer}, {Kellermann}, {Kelley}, {Kheirandish}, {Kin},
  {Kintscher}, {Kiryluk}, {Klein}, {Koirala}, {Kolanoski}, {Kontrimas},
  {K{\"o}pke}, {Kopper}, {Kopper}, {Koskinen}, {Koundal}, {Kovacevich},
  {Kowalski}, {Kozynets}, {Kun}, {Kurahashi}, {Lad}, {Lagunas Gualda},
  {Lanfranchi}, {Larson}, {Lauber}, \& {Lazar}}]{2022Sci...378..538I}
{IceCube Collaboration}, {Abbasi}, R., {Ackermann}, M., {et~al.} 2022, Science,
  378, 538, \dodoi{10.1126/science.abg3395}

\bibitem[{{Icecube Collaboration} {et~al.}(2023){Icecube Collaboration},
  {Abbasi}, {Ackermann}, {Adams}, {Aguilar}, {Ahlers}, {Ahrens}, {Alameddine},
  {Alves}, {Amin}, {Andeen}, {Anderson}, {Anton}, {Arguelles}, {Ashida},
  {Athanasiadou}, {Axani}, {Bai}, {Balagopal}, {Barwick}, {Basu}, {Baur},
  {Bay}, {Beatty}, {Becker}, {Becker Tjus}, {Beise}, {Bellenghi}, {Benda},
  {Benzvi}, {Berley}, {Bernardini}, {Besson}, {Binder}, {Bindig}, {Blaufuss},
  {Blot}, {Boddenberg}, {Bontempo}, {Book}, {Borowka}, {Boser}, {Botner},
  {Bottcher}, {Bourbeau}, {Bradascio}, {Braun}, {Brinson}, {Bron},
  {Brostean-Kaiser}, {Burley}, {Busse}, {Campana}, {Carnie-Bronca}, {Chen},
  {Chen}, {Chirkin}, {Choi}, {Clark}, {Clark}, {Classen}, {Coleman}, {Collin},
  {Connolly}, {Conrad}, {Coppin}, {Correa}, {Cowen}, {Cross}, {Dappen}, {Dave},
  {de Clercq}, {Delaunay}, {Delgado Lopez}, {Dembinski}, {Deoskar}, {Desai},
  {Desiati}, {de Vries}, {de Wasseige}, {Deyoung}, {Diaz}, {Diaz-Velez},
  {Dittmer}, {Dujmovic}, {Dunkman}, {Duvernois}, {Ehrhardt}, {Eller}, {Engel},
  {Erpenbeck}, {Evans}, {Evenson}, {Fan}, {Fazely}, {Fedynitch}, {Feigl},
  {Fiedlschuster}, {Fienberg}, {Finley}, {Fischer}, {Fox}, {Franckowiak},
  {Friedman}, {Fritz}, {Furst}, {Gaisser}, {Gallagher}, {Ganster}, {Garcia},
  {Garrappa}, {Gerhardt}, {Ghadimi}, {Glaser}, {Glauch}, {Glusenkamp},
  {Goehlke}, {Goldschmidt}, {Gonzalez}, {Goswami}, {Grant}, {Gregoire},
  {Griswold}, {Gunther}, {Gutjahr}, {Haack}, {Hallgren}, {Halliday}, {Halve},
  {Halzen}, {Ha}, {Hanson}, {Hardin}, {Harnisch}, {Haungs}, {Helbing},
  {Henningsen}, {Hettinger}, {Hickford}, {Hignight}, {Hill}, {Hill}, {Hoffman},
  {Hoshina}, {Hou}, {Huang}, {Huber}, {Huber}, {Hultqvist}, {Hunnefeld},
  {Hussain}, {Hymon}, {in}, {Iovine}, {Ishihara}, {Jansson}, {Japaridze},
  {Jeong}, {Jin}, {Jones}, {Kang}, {Kang}, {Kang}, {Kappes}, {Kappesser},
  {Kardum}, {Karg}, {Karl}, {Karle}, {Katz}, {Kauer}, {Kellermann}, {Kelley},
  {Kheirandish}, {Kin}, {Kiryluk}, {Klein}, {Kochocki}, {Koirala}, {Kolanoski},
  {Kontrimas}, {Kopke}, {Kopper}, {Kopper}, {Koskinen}, {Koundal},
  {Kovacevich}, {Kowalski}, {Kozynets}, {Krupczak}, {Kun}, {Kurahashi}, {Lad},
  {Lagunas Gualda}, {Lanfranchi}, {Larson}, {Lauber}, {Lazar}, {Lee}, \&
  {Leonard}}]{2023Sci...380.1338I}
{Icecube Collaboration}, {Abbasi}, R., {Ackermann}, M., {et~al.} 2023, Science,
  380, 1338, \dodoi{10.1126/science.adc9818}

\bibitem[{{Inoue} {et~al.}(2021){Inoue}, {Yabe}, \&
  {Ueda}}]{2021PASJ...73.1315I}
{Inoue}, Y., {Yabe}, K., \& {Ueda}, Y. 2021, \pasj, 73, 1315,
  \dodoi{10.1093/pasj/psab077}

\bibitem[{{Jones} {et~al.}(2009){Jones}, {Read}, {Saunders}, {Colless},
  {Jarrett}, {Parker}, {Fairall}, {Mauch}, {Sadler}, {Watson}, {Burton},
  {Campbell}, {Cass}, {Croom}, {Dawe}, {Fiegert}, {Frankcombe}, {Hartley},
  {Huchra}, {James}, {Kirby}, {Lahav}, {Lucey}, {Mamon}, {Moore}, {Peterson},
  {Prior}, {Proust}, {Russell}, {Safouris}, {Wakamatsu}, {Westra}, \&
  {Williams}}]{2009MNRAS.399..683J}
{Jones}, D.~H., {Read}, M.~A., {Saunders}, W., {et~al.} 2009, \mnras, 399, 683,
  \dodoi{10.1111/j.1365-2966.2009.15338.x}

\bibitem[{{Jorstad} {et~al.}(2005){Jorstad}, {Marscher}, {Lister}, {Stirling},
  {Cawthorne}, {Gear}, {G{\'o}mez}, {Stevens}, {Smith}, {Forster}, \&
  {Robson}}]{2005AJ....130.1418J}
{Jorstad}, S.~G., {Marscher}, A.~P., {Lister}, M.~L., {et~al.} 2005, \aj, 130,
  1418, \dodoi{10.1086/444593}

\bibitem[{{Kavanagh} {et~al.}(2015){Kavanagh}, {Sasaki}, {Bozzetto},
  {Filipovi{\'c}}, {Points}, {Maggi}, \& {Haberl}}]{2015A&A...573A..73K}
{Kavanagh}, P.~J., {Sasaki}, M., {Bozzetto}, L.~M., {et~al.} 2015, \aap, 573,
  A73, \dodoi{10.1051/0004-6361/201424354}

\bibitem[{{Kavanagh} {et~al.}(2019){Kavanagh}, {Vink}, {Sasaki}, {Chu},
  {Filipovi{\'c}}, {Ohm}, {Haberl}, {Manojlovic}, \&
  {Maggi}}]{2019A&A...621A.138K}
{Kavanagh}, P.~J., {Vink}, J., {Sasaki}, M., {et~al.} 2019, \aap, 621, A138,
  \dodoi{10.1051/0004-6361/201833659}

\bibitem[{{KM3NeT Collaboration} {et~al.}(2025){KM3NeT Collaboration},
  {MessMapp Group}, {Fermi-LAT Collaboration}, {Owens Valley Radio Observatory
  40-m Telescope Group}, {SVOM Collaboration}, {Baldini}, {Buchner}, {Erkenov},
  {Globus}, {Merloni}, {Paggi}, {Popkov}, {Porquet}, {Salvato}, {Sotnikova}, \&
  {Voitsik}}]{2025arXiv250208484K}
{KM3NeT Collaboration}, {MessMapp Group}, {Fermi-LAT Collaboration}, {et~al.}
  2025, arXiv e-prints, arXiv:2502.08484, \dodoi{10.48550/arXiv.2502.08484}

\bibitem[{{Koribalski} {et~al.}(2004){Koribalski}, {Staveley-Smith}, {Kilborn},
  {Ryder}, {Kraan-Korteweg}, {Ryan-Weber}, {Ekers}, {Jerjen}, {Henning},
  {Putman}, {Zwaan}, {de Blok}, {Calabretta}, {Disney}, {Minchin}, {Bhathal},
  {Boyce}, {Drinkwater}, {Freeman}, {Gibson}, {Green}, {Haynes}, {Juraszek},
  {Kesteven}, {Knezek}, {Mader}, {Marquarding}, {Meyer}, {Mould}, {Oosterloo},
  {O'Brien}, {Price}, {Sadler}, {Schr{\"o}der}, {Stewart}, {Stootman}, {Waugh},
  {Warren}, {Webster}, \& {Wright}}]{Koribalski2004}
{Koribalski}, B.~S., {Staveley-Smith}, L., {Kilborn}, V.~A., {et~al.} 2004,
  \aj, 128, 16, \dodoi{10.1086/421744}

\bibitem[{{Lacy} {et~al.}(2020){Lacy}, {Baum}, {Chandler}, {Chatterjee},
  {Clarke}, {Deustua}, {English}, {Farnes}, {Gaensler}, {Gugliucci},
  {Hallinan}, {Kent}, {Kimball}, {Law}, {Lazio}, {Marvil}, {Mao}, {Medlin},
  {Mooley}, {Murphy}, {Myers}, {Osten}, {Richards}, {Rosolowsky}, {Rudnick},
  {Schinzel}, {Sivakoff}, {Sjouwerman}, {Taylor}, {White}, {Wrobel},
  {Andernach}, {Beasley}, {Berger}, {Bhatnager}, {Birkinshaw}, {Bower},
  {Brandt}, {Brown}, {Burke-Spolaor}, {Butler}, {Comerford}, {Demorest}, {Fu},
  {Giacintucci}, {Golap}, {G{\"u}th}, {Hales}, {Hiriart}, {Hodge}, {Horesh},
  {Ivezi{\'c}}, {Jarvis}, {Kamble}, {Kassim}, {Liu}, {Loinard}, {Lyons},
  {Masters}, {Mezcua}, {Moellenbrock}, {Mroczkowski}, {Nyland}, {O'Dea},
  {O'Sullivan}, {Peters}, {Radford}, {Rao}, {Robnett}, {Salcido}, {Shen},
  {Sobotka}, {Witz}, {Vaccari}, {van Weeren}, {Vargas}, {Williams}, \&
  {Yoon}}]{2020PASP..132c5001L}
{Lacy}, M., {Baum}, S.~A., {Chandler}, C.~J., {et~al.} 2020, \pasp, 132,
  035001, \dodoi{10.1088/1538-3873/ab63eb}

\bibitem[{{Margalit} \& {Quataert}(2021)}]{2021ApJ...923L..14M}
{Margalit}, B., \& {Quataert}, E. 2021, \apjl, 923, L14,
  \dodoi{10.3847/2041-8213/ac3d97}

\bibitem[{{Margalit} \& {Quataert}(2024)}]{MQ24}
---. 2024, \apj, 977, 134, \dodoi{10.3847/1538-4357/ad8b47}

\bibitem[{{Marinelli} {et~al.}(2021){Marinelli}, {Ambrosone}, {Ibnsalih},
  {Miele}, {Migliozzi}, {Pisanti}, {Sharma}, {Ageron}, {Aiello}, {Albert},
  {Alshamsi}, {Alves Garre}, {Aly}, {Ambrosone}, {Ameli}, {Andre},
  {Androulakis}, {Anghinolfi}, {Anguita}, {Anton}, {Ardid}, {Ardid}, {Assal},
  {Aublin}, {Bagatelas}, {Baret}, {Basegmez du Pree}, {Bendahman}, {Benfenati},
  {Berbee}, {van den Berg}, {Bertin}, {Beurthey}, {van Beveren}, {Biagi},
  {Billault}, {Bissinger}, {Boettcher}, {Bou Cabo}, {Boumaaza}, {Bouta},
  {Boutonnet}, {Bouvet}, {Bouwhuis}, {Bozza}, {Br{\^a}nza{\c{s}}}, {Bruijn},
  {Brunner}, {Bruno}, {Buis}, {Buompane}, {Busto}, {Caiffi}, {Caillat},
  {Calvo}, {Campion}, {Capone}, {Carduner}, {Carretero}, {Castaldi}, {Celli},
  {Cereseto}, {Chabab}, {Champion}, {Chau}, {Chen}, {Cherubini}, {Chiarella},
  {Chiarusi}, {Circella}, {Cocimano}, {Coelho}, {Coleiro}, {Colomer Molla},
  {Colonges}, {Coniglione}, {Cosquer}, {Coyle}, {Cresta}, {Creusot}, {Cruz},
  {Cuttone}, {D'Amico}, {Dallier}, {De Martino}, {De Palma}, {Di Palma},
  {D{\'\i}az}, {Diego-Tortosa}, {Distefano}, {Domi}, {Donzaud}, {Dornic},
  {D{\"o}rr}, {Drouhin}, {Eberl}, {Eddyamoui}, {van Eeden}, {van Eijk}, {El
  Bojaddaini}, {Eljarrari}, {Elsaesser}, {Enzenh{\"o}fer}, {Espinosa},
  {Fermani}, {Ferrara}, {Filipovi{\'c}}, {Filippini}, {Fransen}, {Fusco},
  {Gajanana}, {Gal}, {Garc{\'\i}a M{\'e}ndez}, {Garcia Soto}, {Gar{\c{c}}on},
  {Garufi}, {Gatius}, {Gei{\ss}elbrecht}, {Gialanella}, {Giorgio}, {Gozzini},
  {Gracia}, {Graf}, {Grella}, {Guderian}, {Guidi}, {Guillon}, {Guti{\'e}rrez},
  {Haefner}, {Hallmann}, {Hamdaoui}, {van Haren}, {Heijboer}, {Hekalo},
  {Hennig}, {Henry}, {Hern{\'a}ndez-Rey}, {Hofest{\"a}dt}, {Huang}, {Idrissi
  Ibnsalih}, {Ilioni}, {Illuminati}, {James}, {Janezashvili}, {Jansweijer}, {de
  Jong}, {de Jong}, {Jung}, {Kadler}, {Kalaczy{\'n}ski}, {Kalekin}, {Katz},
  {Kayzel}, {Keller}, {Khan Chowdhury}, {Kistauri}, {van der Knaap},
  {Kooijman}, {Kouchner}, {Kreter}, {Kulikovskiy}, {Labalme}, {Lagier},
  {Lahmann}, {Lamare}, {Lamoureux}, {Larosa}, {Lastoria}, {Laurence}, {Lazo},
  {Le Breton}, {Le Guirriec}, {Le Stum}, {Lehaut}, {Leonardi}, {Leone},
  {Leonora}, {Lerouvillois}, {Lesrel}, {Lessing}, {Levi}, {Lincetto}, {Lindsey
  Clark}, {Lipreau}, {Llorens Alvarez}, {Lonardo}, {Longhitano}, {Lopez-Coto},
  {Lumb}, {Maderer}, {Majumdar}, {Ma{\'n}czak}, {Margiotta}, {Marinelli},
  {Marini}, {Markou}, {Martin}, {Mart{\'\i}nez-Mora}, {Martini}, \&
  {Marzaioli}}]{2021JInst..16C2016M}
{Marinelli}, A., {Ambrosone}, A., {Ibnsalih}, W.~I., {et~al.} 2021, Journal of
  Instrumentation, 16, C12016, \dodoi{10.1088/1748-0221/16/12/C12016}

\bibitem[{{Merloni} {et~al.}(2024){Merloni}, {Lamer}, {Liu}, {Ramos-Ceja},
  {Brunner}, {Bulbul}, {Dennerl}, {Doroshenko}, {Freyberg}, {Friedrich},
  {Gatuzz}, {Georgakakis}, {Haberl}, {Igo}, {Kreykenbohm}, {Liu}, {Maitra},
  {Malyali}, {Mayer}, {Nandra}, {Predehl}, {Robrade}, {Salvato}, {Sanders},
  {Stewart}, {Tub{\'\i}n-Arenas}, {Weber}, {Wilms}, {Arcodia}, {Artis},
  {Aschersleben}, {Avakyan}, {Aydar}, {Bahar}, {Balzer}, {Becker}, {Berger},
  {Boller}, {Bornemann}, {Br{\"u}ggen}, {Brusa}, {Buchner}, {Burwitz},
  {Camilloni}, {Clerc}, {Comparat}, {Coutinho}, {Czesla}, {Dannhauer},
  {Dauner}, {Dauser}, {Dietl}, {Dolag}, {Dwelly}, {Egg}, {Ehl}, {Freund},
  {Friedrich}, {Gaida}, {Garrel}, {Ghirardini}, {Gokus}, {Gr{\"u}nwald},
  {Grandis}, {Grotova}, {Gruen}, {Gueguen}, {H{\"a}mmerich}, {Hamaus},
  {Hasinger}, {Haubner}, {Homan}, {Ider Chitham}, {Joseph}, {Joyce},
  {K{\"o}nig}, {Kaltenbrunner}, {Khokhriakova}, {Kink}, {Kirsch}, {Kluge},
  {Knies}, {Krippendorf}, {Krumpe}, {Kurpas}, {Li}, {Liu}, {Locatelli},
  {Lorenz}, {M{\"u}ller}, {Magaudda}, {Mannes}, {McCall}, {Meidinger},
  {Michailidis}, {Migkas}, {Mu{\~n}oz-Giraldo}, {Musiimenta}, {Nguyen-Dang},
  {Ni}, {Olechowska}, {Ota}, {Pacaud}, {Pasini}, {Perinati}, {Pires},
  {Pommranz}, {Ponti}, {Poppenhaeger}, {P{\"u}hlhofer}, {Rau}, {Reh},
  {Reiprich}, {Roster}, {Saeedi}, {Santangelo}, {Sasaki}, {Schmitt},
  {Schneider}, {Schrabback}, {Schuster}, {Schwope}, {Seppi}, {Serim},
  {Shreeram}, {Sokolova-Lapa}, {Starck}, {Stelzer}, {Stierhof}, {Suleimanov},
  {Tenzer}, {Traulsen}, {Tr{\"u}mper}, {Tsuge}, {Urrutia}, {Veronica},
  {Waddell}, {Willer}, {Wolf}, {Yeung}, {Zainab}, {Zangrandi}, {Zhang},
  {Zhang}, \& {Zheng}}]{2024A&A...682A..34M}
{Merloni}, A., {Lamer}, G., {Liu}, T., {et~al.} 2024, \aap, 682, A34,
  \dodoi{10.1051/0004-6361/202347165}

\bibitem[{{Mooley} {et~al.}(2016){Mooley}, {Hallinan}, {Bourke}, {Horesh},
  {Myers}, {Frail}, {Kulkarni}, {Levitan}, {Kasliwal}, {Cenko}, {Cao}, {Bellm},
  \& {Laher}}]{2016ApJ...818..105M}
{Mooley}, K.~P., {Hallinan}, G., {Bourke}, S., {et~al.} 2016, \apj, 818, 105,
  \dodoi{10.3847/0004-637X/818/2/105}

\bibitem[{{Murphy} {et~al.}(2013){Murphy}, {Chatterjee}, {Kaplan}, {Banyer},
  {Bell}, {Bignall}, {Bower}, {Cameron}, {Coward}, {Cordes}, {Croft}, {Curran},
  {Djorgovski}, {Farrell}, {Frail}, {Gaensler}, {Galloway}, {Gendre}, {Green},
  {Hancock}, {Johnston}, {Kamble}, {Law}, {Lazio}, {Lo}, {Macquart}, {Rea},
  {Rebbapragada}, {Reynolds}, {Ryder}, {Schmidt}, {Soria}, {Stairs}, {Tingay},
  {Torkelsson}, {Wagstaff}, {Walker}, {Wayth}, \&
  {Williams}}]{2013PASA...30....6M}
{Murphy}, T., {Chatterjee}, S., {Kaplan}, D.~L., {et~al.} 2013, \pasa, 30,
  e006, \dodoi{10.1017/pasa.2012.006}

\bibitem[{{Murphy} {et~al.}(2021){Murphy}, {Kaplan}, {Stewart}, {O'Brien},
  {Lenc}, {Pintaldi}, {Pritchard}, {Dobie}, {Fox}, {Leung}, {An}, {Bell},
  {Broderick}, {Chatterjee}, {Dai}, {d'Antonio}, {Doyle}, {Gaensler}, {Heald},
  {Horesh}, {Jones}, {McConnell}, {Moss}, {Raja}, {Ramsay}, {Ryder}, {Sadler},
  {Sivakoff}, {Wang}, {Wang}, {Wheatland}, {Whiting}, {Allison}, {Anderson},
  {Ball}, {Bannister}, {Bock}, {Bolton}, {Bunton}, {Chekkala}, {Chippendale},
  {Cooray}, {Gupta}, {Hayman}, {Jeganathan}, {Koribalski}, {Lee-Waddell},
  {Mahony}, {Marvil}, {McClure-Griffiths}, {Mirtschin}, {Ng}, {Pearce},
  {Phillips}, \& {Voronkov}}]{2021PASA...38...54M}
{Murphy}, T., {Kaplan}, D.~L., {Stewart}, A.~J., {et~al.} 2021, \pasa, 38,
  e054, \dodoi{10.1017/pasa.2021.44}

\bibitem[{{Norris} {et~al.}(2021){Norris}, {Marvil}, {Collier}, {Kapi{\'n}ska},
  {O'Brien}, {Rudnick}, {Andernach}, {Asorey}, {Brown}, {Br{\"u}ggen},
  {Crawford}, {English}, {Rahman}, {Filipovi{\'c}}, {Gordon}, {G{\"u}rkan},
  {Hale}, {Hopkins}, {Huynh}, {HyeongHan}, {James Jee}, {Koribalski}, {Lenc},
  {Luken}, {Parkinson}, {Prandoni}, {Raja}, {Reiprich}, {Riseley}, {Shabala},
  {Sheil}, {Vernstrom}, {Whiting}, {Allison}, {Anderson}, {Ball}, {Bell},
  {Bunton}, {Galvin}, {Gupta}, {Hotan}, {Jacka}, {Macgregor}, {Mahony}, {Maio},
  {Moss}, {Pandey-Pommier}, \& {Voronkov}}]{Norris2021}
{Norris}, R.~P., {Marvil}, J., {Collier}, J.~D., {et~al.} 2021, \pasa, 38, e046

\bibitem[{{Padovani} {et~al.}(2024){Padovani}, {Resconi}, {Ajello},
  {Bellenghi}, {Bianchi}, {Blasi}, {Huang}, {Gabici}, {G{\'a}mez Rosas},
  {Niederhausen}, {Peretti}, {Eichmann}, {Guetta}, {Lamastra}, \&
  {Shimizu}}]{2024NatAs...8.1077P}
{Padovani}, P., {Resconi}, E., {Ajello}, M., {et~al.} 2024, Nature Astronomy,
  8, 1077, \dodoi{10.1038/s41550-024-02339-z}

\bibitem[{{Pennock} {et~al.}(2021){Pennock}, {van Loon}, {Filipovi{\'c}},
  {Andernach}, {Haberl}, {Kothes}, {Lenc}, {Rudnick}, {White}, {Agliozzo},
  {Ant{\'o}n}, {Boji{\v{c}}i{\'c}}, {Bomans}, {Collier}, {Crawford}, {Hopkins},
  {Jeganathan}, {Kavanagh}, {Koribalski}, {Leahy}, {Maggi}, {Maitra}, {Marvil},
  {Micha{\l}owski}, {Norris}, {Oliveira}, {Payne}, {Sano}, {Sasaki},
  {Staveley-Smith}, \& {Vardoulaki}}]{2021MNRAS.506.3540P}
{Pennock}, C.~M., {van Loon}, J.~T., {Filipovi{\'c}}, M.~D., {et~al.} 2021,
  MNRAS, 506, 3540, \dodoi{10.1093/mnras/stab1858}

\bibitem[{{Pushkarev} {et~al.}(2023){Pushkarev}, {Aller}, {Aller}, {Homan},
  {Kovalev}, {Lister}, {Pashchenko}, {Savolainen}, \&
  {Zobnina}}]{2023MNRAS.520.6053P}
{Pushkarev}, A.~B., {Aller}, H.~D., {Aller}, M.~F., {et~al.} 2023, \mnras, 520,
  6053, \dodoi{10.1093/mnras/stad525}

\bibitem[{{Rhodes} {et~al.}(2022){Rhodes}, {Fender}, {Motta}, {van den
  Eijnden}, {Williams}, {Bright}, \& {Sivakoff}}]{2022MNRAS.513.2708R}
{Rhodes}, L., {Fender}, R.~P., {Motta}, S., {et~al.} 2022, \mnras, 513, 2708,
  \dodoi{10.1093/mnras/stac954}

\bibitem[{{Rodrigues} {et~al.}(2021){Rodrigues}, {Heinze}, {Palladino}, {van
  Vliet}, \& {Winter}}]{2021PhRvL.126s1101R}
{Rodrigues}, X., {Heinze}, J., {Palladino}, A., {van Vliet}, A., \& {Winter},
  W. 2021, \prl, 126, 191101, \dodoi{10.1103/PhysRevLett.126.191101}

\bibitem[{{Ross} {et~al.}(2021){Ross}, {Callingham}, {Hurley-Walker},
  {Seymour}, {Hancock}, {Franzen}, {Morgan}, {White}, {Bell}, \&
  {Patil}}]{2021MNRAS.501.6139R}
{Ross}, K., {Callingham}, J.~R., {Hurley-Walker}, N., {et~al.} 2021, \mnras,
  501, 6139, \dodoi{10.1093/mnras/staa3795}

\bibitem[{{Ross} {et~al.}(2024){Ross}, {Hurley-Walker}, {Galvin}, {Venville},
  {Duchesne}, {Morgan}, {An}, {G{\"u}rkan}, {Hancock}, {Heald},
  {Johnston-Hollitt}, \& {White}}]{2024PASA...41...54R}
{Ross}, K., {Hurley-Walker}, N., {Galvin}, T.~J., {et~al.} 2024, \pasa, 41,
  e054, \dodoi{10.1017/pasa.2024.57}

\bibitem[{{Sano} {et~al.}(2017){Sano}, {Yamane}, {Voisin}, {Fujii}, {Yoshiike},
  {Inaba}, {Tsuge}, {Babazaki}, {Mitsuishi}, {Yang}, {Aharonian}, {Rowell},
  {Filipovi{\'c}}, {Mizuno}, {Tachihara}, {Kawamura}, {Onishi}, \&
  {Fukui}}]{2017ApJ...843...61S}
{Sano}, H., {Yamane}, Y., {Voisin}, F., {et~al.} 2017, \apj, 843, 61,
  \dodoi{10.3847/1538-4357/aa73e0}

\bibitem[{{Sault} {et~al.}(1995){Sault}, {Teuben}, \& {Wright}}]{Sault1995}
{Sault}, R.~J., {Teuben}, P.~J., \& {Wright}, M.~C.~H. 1995, in Astronomical
  Society of the Pacific Conference Series, Vol.~77, Astronomical Data Analysis
  Software and Systems IV, ed. R.~A. {Shaw}, H.~E. {Payne}, \& J.~J.~E.
  {Hayes}, 433, \dodoi{10.48550/arXiv.astro-ph/0612759}

\bibitem[{{Smeaton} {et~al.}(2024){Smeaton}, {Filipovi{\'c}}, {Lazarevi{\'c}},
  {Alsaberi}, {Ahmad}, {Araya}, {Ball}, {Bordiu}, {Buemi}, {Bufano}, {Dai},
  {Haberl}, {Hopkins}, {Ingallinera}, {Jarrett}, {Koribalski}, {Kothes},
  {Kraan-Korteweg}, {Leahy}, {Lundqvist}, {Maitra}, {Martin}, {Payne},
  {Rowell}, {Sano}, {Sasaki}, {Soria}, {Steyn}, {Umana}, {Uro{\v{s}}evi{\'c}},
  {Velovi{\'c}}, {Vernstrom}, {Vukoti{\'c}}, \& {West}}]{2024MNRAS.534.2918S}
{Smeaton}, Z.~J., {Filipovi{\'c}}, M.~D., {Lazarevi{\'c}}, S., {et~al.} 2024,
  \mnras, 534, 2918, \dodoi{10.1093/mnras/stae2237}

\bibitem[{{Sommani} {et~al.}(2024){Sommani}, {Franckowiak}, {Lincetto}, \&
  {Dettmar}}]{2024arXiv240303752S}
{Sommani}, G., {Franckowiak}, A., {Lincetto}, M., \& {Dettmar}, R.-J. 2024,
  arXiv e-prints, arXiv:2403.03752, \dodoi{10.48550/arXiv.2403.03752}

\bibitem[{{Soria} {et~al.}(2010){Soria}, {Pakull}, {Broderick}, {Corbel}, \&
  {Motch}}]{2010MNRAS.409..541S}
{Soria}, R., {Pakull}, M.~W., {Broderick}, J.~W., {Corbel}, S., \& {Motch}, C.
  2010, \mnras, 409, 541, \dodoi{10.1111/j.1365-2966.2010.17360.x}

\bibitem[{Stewart {et~al.}(2024{\natexlab{a}})Stewart, Dobie, O'Brien, \&
  Kaplan}]{adam_stewart_2024_13131754}
Stewart, A., Dobie, D., O'Brien, A., \& Kaplan, D. 2024{\natexlab{a}},
  askap-vast/vast-tools: v3.1.1, v3.1.1,  Zenodo,
  \dodoi{10.5281/zenodo.8365236}

\bibitem[{Stewart {et~al.}(2024{\natexlab{b}})Stewart, Serg, O'Brien, Liptai,
  Dobie, Mauch, Saleheen, Wang, Pritchard, \&
  Jazzy}]{adam_stewart_2024_14048598}
Stewart, A., Serg, O'Brien, A., {et~al.} 2024{\natexlab{b}},
  askap-vast/vast-pipeline: v1.2.0, v1.2.0,  Zenodo,
  \dodoi{10.5281/zenodo.13927015}

\bibitem[{{The KM3NeT collaboration}(2025)}]{2025arXiv250208508T}
{The KM3NeT collaboration}. 2025, arXiv e-prints, arXiv:2502.08508,
  \dodoi{10.48550/arXiv.2502.08508}

\bibitem[{{The KM3NeT Collaboration} {et~al.}(2025){The KM3NeT Collaboration},
  {Aiello}, {Albert}, {Alhebsi}, {Alshamsi}, {Alves Garre}, {Ambrosone},
  {Ameli}, {Andre}, {Anghinolfi}, {Aphecetche}, {Ardid}, {Ardid},
  {Arg{\"u}elles}, {Atmani}, {Aublin}, {Badaracco}, {Bailly-Salins},
  {Barda{\v{c}}ov{\'a}}, {Baret}, {Bariego-Quintana}, {Becherini}, {Bendahman},
  {Benfenati Gualandi}, {Benhassi}, {Bennani}, {Benoit}, {Berbee}, {Bertin},
  {Biagi}, {Boettcher}, {Bonanno}, {Bouasla}, {Boumaaza}, {Bouta}, {Bouwhuis},
  {Bozza}, {Bozza}, {Br{\^a}nza{\c{s}}}, {Bretaudeau}, {Breuhaus}, {Bruijn},
  {Brunner}, {Bruno}, {Buis}, {Buompane}, {Buson}, {Busto}, {Caiffi}, {Calvo},
  {Capone}, {Carenini}, {Carretero}, {Cartraud}, {Castaldi}, {Cecchini},
  {Celli}, {Cerisy}, {Chabab}, {Chen}, {Cherubini}, {Chiarusi}, {Circella},
  {Cocimano}, {Coelho}, {Coleiro}, {Colonges}, {Condorelli}, {Coniglione},
  {Coyle}, {Creusot}, {Cuttone}, {D'Amico}, {Dallier}, {De Benedittis}, {De
  Martino}, {De Wasseige}, {Decoene}, {Del Rosso}, {Di Mauro}, {Di Palma},
  {Diaz}, {Diego-Tortosa}, {Distefano}, {Domi}, {Donzaud}, {Dornic},
  {Drakopoulou}, {Drouhin}, {Ducoin}, {Dvornick{\'y}}, {Eberl}, {Eckerov{\'a}},
  {Eddymaoui}, {van Eeden}, {Eff}, {van Eijk}, {El Bojaddaini}, {El Hedri},
  {Ellajosyula}, {Enzenh{\"o}fer}, {Ferrara}, {Filipovi{\'c}v}, {Filippini},
  {Franciotti}, {Fusco}, {Gagliardini}, {Gal}, {Garc{\'\i}a M{\'e}ndez},
  {Garcia Soto}, {Gatius Oliver}, {Gei{\ss}elbrecht}, {Genton}, {Ghaddari},
  {Gialanella}, {Gibson}, {Giorgio}, {Goos}, {Goswami}, {Gozzini}, {Gracia},
  {Graf}, {Guidi}, {Guillon}, {Guti{\'e}rrez}, {Haack}, {van Haren},
  {Heijboer}, {Hennig}, {Henry}, {Hern{\'a}ndez-Rey}, {Idrissi Ibnsalih},
  {Ilioni}, {Illuminati}, {Joly}, {de Jong}, {de Jong}, {Jung},
  {Kalaczy{\'n}ski}, {Kalekin}, {Kamp}, {Katz}, {Kistauri}, {Kopper},
  {Kouchner}, {Kovalev}, {Kueviakoe}, {Kulikovskiy}, {Kvatadze}, {Labalme},
  {Lahmann}, {Lamoureux}, {Lancelin}, {Larosa}, {Lastoria}, {Lazar}, {Lazo},
  {Le Stum}, {Lehaut}, {Lemaitre}, {Leonora}, {Lessing}, {Levi}, {Lincetto},
  {Lindsey Clark}, {Longhitano}, {Lumb}, {Magnani}, {Majumdar}, {Malerba},
  {Mamedov}, {Manfreda}, {Marconi}, {Margiotta}, {Marinelli}, {Markou},
  {Martin}, {Marzaioli}, {Mastrodicasa}, {Mastroianni}, {Mauro}, {Miele},
  {Migliozzi}, {Migneco}, {Mitsou}, {Mollo}, {Mongelli}, {Morales-Gallegos},
  {Moussa}, {Mozun Mateo}, {Muller}, {Musone}, {Musumeci}, {Navas},
  {Nayerhoda}, {Nicolau}, {Nkosi}, {Fearraigh}, {Oliviero}, \&
  {Orlando}}]{KM3nat}
{The KM3NeT Collaboration}, {Aiello}, S., {Albert}, A., {et~al.} 2025, \nat,
  638, 376, \dodoi{10.1038/s41586-024-08543-1}

\bibitem[{{Velovi{\'c}} {et~al.}(2022){Velovi{\'c}}, {Filipovi{\'c}}, {Barnes},
  {Norris}, {Tremblay}, {Heald}, {Rudnick}, {Shabala}, {Pannuti}, {Andernach},
  {Titov}, {Waddell}, {Koribalski}, {Grupe}, {Jarrett}, {Alsaberi}, {Carretti},
  {Collier}, {Einecke}, {Galvin}, {Hotan}, {Manojlovi{\'c}}, {Marvil},
  {Nandra}, {Reiprich}, {Rowell}, {Salvato}, \&
  {Whiting}}]{2022MNRAS.516.1865V}
{Velovi{\'c}}, V., {Filipovi{\'c}}, M.~D., {Barnes}, L., {et~al.} 2022, \mnras,
  516, 1865, \dodoi{10.1093/mnras/stac2012}

\bibitem[{{Vieu} \& {Reville}(2023)}]{2023MNRAS.519..136V}
{Vieu}, T., \& {Reville}, B. 2023, \mnras, 519, 136,
  \dodoi{10.1093/mnras/stac3469}

\bibitem[{{Walton} {et~al.}(2022){Walton}, {Mackenzie}, {Gully}, {Patel},
  {Roberts}, {Earnshaw}, \& {Mateos}}]{2022MNRAS.509.1587W}
{Walton}, D.~J., {Mackenzie}, A.~D.~A., {Gully}, H., {et~al.} 2022, \mnras,
  509, 1587, \dodoi{10.1093/mnras/stab3001}

\bibitem[{{Wang} {et~al.}(2016){Wang}, {Koribalski}, {Serra}, {van der Hulst},
  {Roychowdhury}, {Kamphuis}, \& {Chengalur}}]{Wang2016}
{Wang}, J., {Koribalski}, B.~S., {Serra}, P., {et~al.} 2016, \mnras, 460, 2143,
  \dodoi{10.1093/mnras/stw1099}

\bibitem[{{Wong} {et~al.}(2025){Wong}, {Garon}, {Alger}, {Rudnick}, {Shabala},
  {Willett}, {Banfield}, {Andernach}, {Norris}, {Swan}, {Hardcastle},
  {Lintott}, {White}, {Seymour}, {Kapi{\'n}ska}, {Tang}, {Simmons}, \&
  {Schawinski}}]{2025MNRAS.536.3488W}
{Wong}, O.~I., {Garon}, A.~F., {Alger}, M.~J., {et~al.} 2025, \mnras, 536,
  3488, \dodoi{10.1093/mnras/stae2790}

\bibitem[{{Wright} {et~al.}(2010){Wright}, {Eisenhardt}, {Mainzer}, {Ressler},
  {Cutri}, {Jarrett}, {Kirkpatrick}, {Padgett}, {McMillan}, {Skrutskie},
  {Stanford}, {Cohen}, {Walker}, {Mather}, {Leisawitz}, {Gautier}, {McLean},
  {Benford}, {Lonsdale}, {Blain}, {Mendez}, {Irace}, {Duval}, {Liu}, {Royer},
  {Heinrichsen}, {Howard}, {Shannon}, {Kendall}, {Walsh}, {Larsen}, {Cardon},
  {Schick}, {Schwalm}, {Abid}, {Fabinsky}, {Naes}, \&
  {Tsai}}]{2010AJ....140.1868W}
{Wright}, E.~L., {Eisenhardt}, P. R.~M., {Mainzer}, A.~K., {et~al.} 2010, \aj,
  140, 1868, \dodoi{10.1088/0004-6256/140/6/1868}

\bibitem[{{Yamane} {et~al.}(2021){Yamane}, {Sano}, {Filipovi{\'c}}, {Tokuda},
  {Fujii}, {Babazaki}, {Mitsuishi}, {Inoue}, {Aharonian}, {Inaba}, {Inutsuka},
  {Maxted}, {Mizuno}, {Onishi}, {Rowell}, {Tsuge}, {Voisin}, {Yoshiike},
  {Fukuda}, {Kawamura}, {Bamba}, {Tachihara}, \& {Fukui}}]{2021ApJ...918...36Y}
{Yamane}, Y., {Sano}, H., {Filipovi{\'c}}, M.~D., {et~al.} 2021, \apj, 918, 36,
  \dodoi{10.3847/1538-4357/ac0adb}

\bibitem[{{Yasuda} {et~al.}(2001){Yasuda}, {Fukugita}, {Narayanan}, {Lupton},
  {Strateva}, {Strauss}, {Ivezi{\'c}}, {Kim}, {Hogg}, {Weinberg}, {Shimasaku},
  {Loveday}, {Annis}, {Bahcall}, {Blanton}, {Brinkmann}, {Brunner}, {Connolly},
  {Csabai}, {Doi}, {Hamabe}, {Ichikawa}, {Ichikawa}, {Johnston}, {Knapp},
  {Kunszt}, {Lamb}, {McKay}, {Munn}, {Nichol}, {Okamura}, {Schneider},
  {Szokoly}, {Vogeley}, {Watanabe}, \& {York}}]{2001AJ....122.1104Y}
{Yasuda}, N., {Fukugita}, M., {Narayanan}, V.~K., {et~al.} 2001, \aj, 122,
  1104, \dodoi{10.1086/322093}

\end{thebibliography}

\appendix 

\section{Data}
 \label{sec:data}

\subsection{ASKAP EMU+POSSUM}
 \label{subsec:Emu_Data}

The 30\,deg$^2$ \ac{ASKAP} tile that includes the \km\ event location was observed as part of \ac{EMU} \citep[Project Code AS201;][Hopkins et al., submitted]{Norris2021} and \ac{POSSUM} \citep[Project Code AS203,][Gaensler et al., submitted]{Gaensler2010}. Two \ac{ASKAP} scheduling blocks (SB~59692; tile EMU$\_0626-09$A and SB~61077; EMU$\_0626-09$B) cover this area and were observed on the 2$^{\rm nd}$ of March 2024 and the 13$^{\rm th}$ of April~2024, respectively, i.e. approximately 1~year after the \km\ event. Both tiles use \ac{ASKAP}'s 36 antenna array with a central frequency of 943.5\,MHz, a bandwidth of 288\,MHz, and a 300-minute exposure time. Only SB~59692 was used for source finding due to significant interference from a nearby source in SB~61077. However, SB~61077 was still able to be used for polarimetric analysis. The high-resolution \ac{EMU} image (B.S.=10\arcsec) of SB~59692 was also used to generate the spectral index map for UGCA~127 (see Sec.~\ref{subsubsec:ugc_Discussion}) as the inner components could not be resolved in the convolved 15\arcsec\ image. 

The data were processed using the standard ASKAPSoft pipelines~\citep{2019ascl.soft12003G}, producing both a multifrequency synthesis (MFS) band-averaged Stokes~$I$ image for \ac{EMU}, and full Stokes~$I$, $Q$, $U$ and $V$ frequency cubes with 1\,MHz channels for \ac{POSSUM}. De-rotation of Stokes~$Q$ and $U$ were used to create Faraday spectra, with the amplitude of the peak in the spectrum taken as the polarized intensity. In Figure~\ref{fig:fig1} we show the convolved EMU Stokes~$I$ radio continuum image which has a resolution of 15\arcsec\ and an \ac{RMS} noise sensitivity of $\sigma$\,=\,50\,$\mu$Jy\,beam$^{-1}$. We find no Stokes~$V$ emission within the search area. To obtain Stokes~$Q$, $U$, polarized intensity (PI) and polarization angle (PA) images, we utilized the \ac{RM} synthesis technique~\citep{2010ApJ...712.1157H}. We did not use the Fourier transform method from the POSSUM pipeline but the de-rotation technique as described in \citet{Ball2023}.

\subsection{Other radio catalogs}
\label{subsec:Catalog_Data}

We use several previous radio catalogs and surveys which overlap the search area to cross-match our sources and include the flux measurements in our catalog to calculate spectral indices (see Table~\ref{Appendix:Sample}). We use data from the \ac{GLEAM-X} survey \citep[][$\nu$\,=\,72$-$231\,MHz; B.S.$\sim$ 45\arcsec]{2024PASA...41...54R}, the \ac{RACS} \citep[][$\nu$\,=\,888\,MHz; B.S.\,=\,25\arcsec]{2021PASA...38...58H}, the \ac{NVSS} \citep[][$\nu$\,=\,1.4\,GHz; B.S.\,=\,45\arcsec]{1998AJ....115.1693C}, and the \ac{VLASS} Quick-Look epoch-averaged images \citep[($\nu$\,=\,3\,GHz; B.S.\,$\sim$\,3\arcsec)]{2021ApJS..255...30G}. Limitations for the \ac{VLASS} Quick-Look images are discussed in \citet{2021ApJS..255...30G}. We also use the different epochs in the \ac{VLASS} Quick-Look and \ac{ASKAP} \ac{VAST} surveys~\citep{2013PASA...30....6M, 2021PASA...38...54M} to search for variability.

As the \ac{ASKAP} \ac{EMU} tile does not cover the entire 68\% confidence region, we only catalog sources that appear within the field of view. The entire region is analyzed with the \ac{VLASS} data.

\clearpage

\section{Example Table}
\label{Appendix:Sample}

    \begin{table}[b]

        \rotatebox{90}{%
        \begin{minipage}{9cm}
        \vskip3cm
        \centering
        \resizebox{2.2\textwidth}{!}{%
        \hskip-6cm\begin{tabular}{lccccccccccccc}
        \hline
        [1] & [2] & [3] & [4] & [5] & [9] & [10] & [11] & [12] & [13] & [14] & [15] \\
        NAME(S) & RA (J2000) & DEC (J2000) & EMU & GLEAM W087 & GLEAM W215 & RACS & NVSS & VLASS & $\alpha$ & n$^\alpha$ & $z$ \\
        &  &  & 944 (MHz) & 87 (MHz) & 215 (MHz) & 888 (MHz) & 1.40 (GHz) & 3.00 (GHz) &  &  &  \\
        &  &  & $S\pm\Delta S$ & $S\pm\Delta S$ & $S\pm\Delta S$ & $S\pm\Delta S$ & $S\pm\Delta S$ & $S\pm\Delta S$ & $\alpha\pm\Delta\alpha$ &  & $z\pm\Delta z$ &  \\
        & hh:mm:ss.ss & dd:mm:ss.ss & (Jy) & (Jy) & (Jy) & (Jy) & (Jy) & (Jy) &  &  &  \\
        \hline
        $^{a}$EMU J062056--082949 & 06:20:56.50 & -08:29:49.80 & 0.255$\pm$0.026 & 1.017$\pm$0.102 & 0.559$\pm$0.056 & -- & -- & --
        & -0.61$\pm$0.01 & 7 & -- \\
        UGCA 127 & & & & & & & & & & & \\
        Phaedra (\name) & & & & & & & & & & & \\
        $^a$EMU J061716--075449 & 06:17:16.50 & -07:54:49.60 & 0.045$\pm$0.005 & 0.216$\pm$0.022 & 0.117$\pm$0.012 & -- & -- & --
        & -0.65$\pm$0.02 & 7 & 0.13$\pm$0.01 \\
        \wisegala & & & & & & & & & & & \\
        Hebe (\nametwo) & & & & & & & & & & & \\
        $^a$EMU J062244--072333 & 06:22:44.90 & -07:23:33.40 & 0.122$\pm$0.012 & 0.019$\pm$-0.002 & 0.003$\pm$0.0003 & 0.110$\pm$0.001 & 0.106$\pm$0.003 & 0.0828$\pm$0.0046 & 1.09$\pm$0.21 & 8 & -- \\
        $^a$EMU J062248--072246 & 06:22:48.40 & -07:22:46.20 & 0.002$\pm$0.0002 & -- & -- & -- & -- & 0.0034$\pm$0.0047 & -- & -- & -- \\
        EMU J062053--084851 & 06:20:53.40 & -08:48:51.70 & 0.106$\pm$0.011 & -- & -- & -- & -- & 0.0024$\pm$0.0007 & -- & -- & -- \\
        EMU J062039--073935 & 06:20:39.90 & -07:39:35.60 & 0.129$\pm$0.013 & 0.547$\pm$0.055 & 0.306$\pm$0.031 & -- & -- & 0.0079$\pm$0.0013 & -1.10$\pm$0.16 & 7 & 0.13$\pm$0.01 \\
        EMU J061946--064201 & 06:19:46.40 & -06:42:01.10 & 0.009$\pm$0.001 & 0.176$\pm$0.018 & 0.066$\pm$0.007 & -- & -- & 0.0011$\pm$0.0005 & -1.46$\pm$0.08 & 7 & -- \\
        EMU J061944--064232 & 06:19:44.50 & -06:42:32.70 & 0.011$\pm$0.001 & -- & -- & -- & -- & 0.0006$\pm$0.0005 & -- & -- & -- \\
        EMU J061542--063844 & 06:15:42.00 & -06:38:44.50 & 0.108$\pm$0.011 & 0.547$\pm$0.055 & 0.229$\pm$0.023 & -- & -- & 0.0017$\pm$0.0010 & -1.45$\pm$0.26 & 7 & -- \\
        EMU J061931--063426 & 06:19:31.00 & -06:34:26.50 & 0.002$\pm$0.0002 & -- & -- & -- & 0.003$\pm$0.0005 & 0.0024$\pm$0.0003 & -0.02$\pm$0.14 & 3 & -- \\
        EMU J062118--073729 & 06:21:18.20 & -07:37:29.30 & 0.007$\pm$0.001 & -- & -- & -- & 0.004$\pm$0.0004 & 0.0032$\pm$0.0004 & -0.65$\pm$0.29 & 3 & -- \\
        EMU J061626--070526 & 06:16:26.80 & -07:05:26.90 & 0.004$\pm$0.0004 & -- & -- & 0.003$\pm$0.0008 & -- & 0.0019$\pm$0.0004 & -0.44$\pm$0.32 & 3 & -- \\
        EMU J061345--071315 & 06:13:45.10 & -07:13:15.80 & 0.002$\pm$0.0003 & -- & -- & 0.003$\pm$0.0007 & -- & 0.0011$\pm$0.0003 & -0.76$\pm$0.11 & 3 & -- \\
        EMU J061438--071836 & 06:14:38.60 & -07:18:36.50 & 0.002$\pm$0.0002 & -- & -- & 0.003$\pm$0.001 & -- & 0.0011$\pm$0.0004 & -0.64$\pm$0.41 & 3 & -- \\
        EMU J061346--082442 & 06:13:46.90 & -08:24:42.80 & 0.002$\pm$0.0002 & -- & -- & 0.003$\pm$0.001 & -- & 0.0014$\pm$0.0004 & -0.51$\pm$0.45 & 3 & -- \\
        EMU J061356--080149 & 06:13:56.50 & -08:01:49.40 & 0.004$\pm$0.0004 & -- & -- & 0.004$\pm$0.001 & -- & 0.0016$\pm$0.0004 & -0.71$\pm$0.10 & 3 & -- \\
        EMU J061336--081344 & 06:13:36.90 & -08:13:44.00 & 0.003$\pm$0.0003 & -- & -- & 0.004$\pm$0.001 & -- & 0.0019$\pm$0.0003 & -0.42$\pm$0.17 & 3 & -- \\
        \hline
        \end{tabular}}%
        \vskip.25cm
        \parbox{2\textwidth}{\caption{Example table of the final \ac{EMU} catalog generated as a part of this survey. Column [1] is the EMU designation name, or additional names where applicable. Columns [2] and [3] are FK5 (J2000) RA and DEC positions for a given source. Columns from [4] to [13] are measured fluxes and their respective errors in Janskys (Jy), columns [6], [7] and [8] are not included, but are the central three GLEAM bands; W118, W154, and W185. Column [13] includes calculated spectral indices, and column [14] is the number of points used in spectral index calculations. Column [15] is redshift and its associated error, not included is column [16], which contains the bibliographical reference for provided redshifts. Also not included, column [17] indicates whether a source is extended or not. It is important to note that each column with associated errors is split into two columns in the full catalog. $^a$Sources discussed in this paper.}}
        \label{table:sample_cat}
        \end{minipage}%
        }
    \end{table}   

\clearpage

\section{\HI\ analysis of the galaxy UGCA~127 (Phaedra)}
 \label{App2:HIfig}

\begin{figure}[h!]
\centering  \includegraphics[width=1\linewidth]{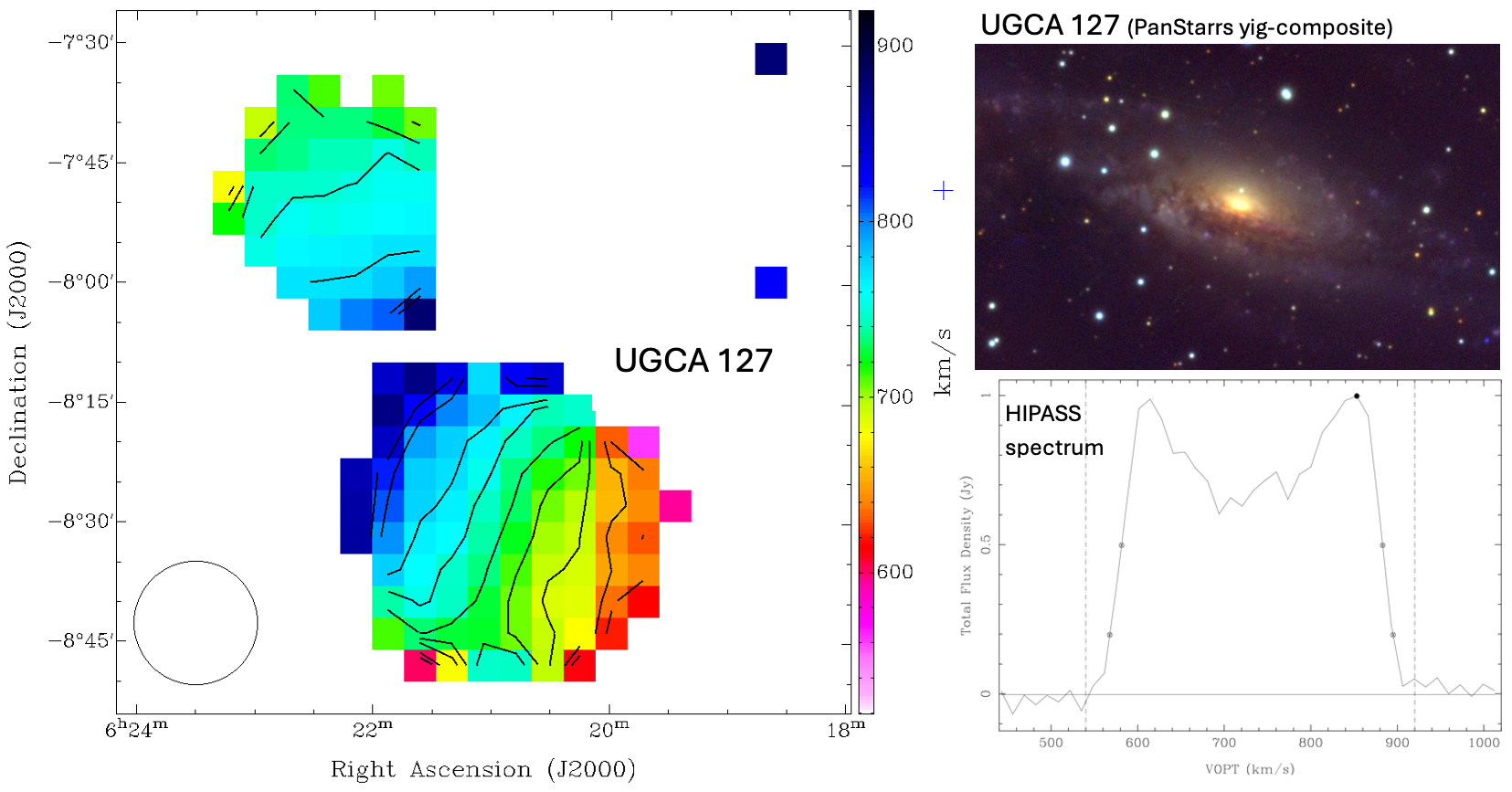}
\caption{{\bf -- Left:} Mean \HI\ velocity field of the galaxy \ugca\ and its companion using data from the \HI\ Parkes All Sky Survey (HIPASS). The Parkes gridded beam (15.5\arcmin) is shown in the bottom left corner. {\bf -- Top right:} PanSTARRS optical color-composite image using the $yig$ bands. {\bf -- Bottom right:} Integrated HIPASS spectrum; the 50\% and 20\% velocity width estimates as well as the peak flux density are indicated. 
}
\label{fig:figHI}
\end{figure}

UGCA~127 (HIPASS J0620--08) shows a symmetric double-horn \HI\ profile (Figure~\ref{fig:figHI}), with a \HI\ flux density of 250\,Jy\,km\,s$^{-1}$ and a \HI\ mass of $4\times10^9$~$M_{\odot}$. 
We estimate a rotational velocity of 187~km\,s$^{-1}$, suggesting a total dynamical mass of $M_{\rm dyn}$ = $1.4 \times 10^{11}$~$M_{\odot}$ for an \HI\ disk radius of 17\,kpc \citep[using the \HI\ mass -- size relation,][]{Wang2016}. A gas-rich companion (HIPASS~J0622--07) is detected $\sim$45\arcmin\ to the north-east. \citet{2021PASJ...73.1315I} estimate a \ac{SFR} of 0.36~$M_{\odot}$\,yr$^{-1}$ and a stellar mass of $2 \times 10^{10}~M_{\odot}$.

\end{document}